# Stochastic quantum hydrodynamic model from the dark matter of vacuum fluctuations: The Langevin-Schrödinger equation and the large-scale classical limit

Simone Chiarelli[1] and Piero Chiarelli[2,3]

*(1) Scuola Normale Superiore, Consoli del Mare, 1 , 56126, Pisa, Italy*

*Email: simone.chiarelli@sns.it*
Phone: +39-050-509686
Fax: +39-050.563513

*(2) National Council of Research of Italy, San Cataldo, Moruzzi 1, 56124, Pisa, Italy*

Email: pchiare@ifc.cnr.it.
Phone: +39-050-315-2359
Fax: +39-050-315-2166

*(3) Interdepartmental Center "E.Piaggio", Faculty of Engineering, University of Pisa, Diotisalvi 2, 56122, Pisa, Italy.*

Abstract: The work derives the quantum evolution in a fluctuating vacuum by introducing the related (dark) mass density noise into the Madelung quantum hydrodynamic model. The paper shows that the classical dynamics can spontaneously emerge on the cosmological scale allowing the realization of the classical system-environment super system. The work shows that the dark matter-induced noise is not spatially white and owns a well defined correlation function with the intrinsic vacuum physical length given by the De Broglie one. The resulting model, in the case of microscopic systems, whose dimension is much smaller than the De Broglie length, leads to the Langevin-Schrodinger equation whose friction coefficient is not constant. The derivation puts in evidence the range of application of the Langevin-Schrodinger equation and the approximations inherent to its foundation.

The work shows that the classical physics can be achieved in a description whose length scale is much bigger both than the De Broglie length and the quantum potential range of interaction. The model shows that the quantum-to-classical transition is not possible in linear systems, and defines the long-distance characteristics as well as the range of interaction of the non-local quantum potential in order to have a coarse-grained large-scale classical phase. The theory also shows that the process of measurement (by a large-scale classical observer) satisfies the minimum uncertainty conditions if interactions and information do not travel faster than the light speed, reconciling the quantum entanglement with the relativistic macroscopic locality.

## 1. Introduction

The conflict between the quantum mechanics and the classical ones attracts the interest of many researchers of the noways physics [1-3].
This lack of knowledge has lead to many logical paradoxes that contrast with our sense of reality [1-3]. A quantitative tentative to investigate the problem was given by Bell [3] in response to the so called EPR paradox [2] a critical analysis of the quantum non-locality respect to the notion of the macroscopic *classical freedom* and *local relativistic causality*.
The Copenhagen interpretation of quantum mechanics [3–5] treats the wave function as representing 'the probability' of finding a particle at some location' [12]. However, such a treatment leads to the non-intuitive conclusion that the physical state is just a probability wave until observed. The absence of an analytical link with the pre-measure physical state fights against the common sense of reality and the existence of a real world independent by the observer and the measure process [5].
If the Copenhagen probabilistic connection with the pre-measure world is strictly assumed, the conclusion that the real state is not physically defined before the measure is unavoidable.
Actually, the completeness and self-consistency of this logical result cannot be achieved since the process of the observation is out the Hamiltonian description of quantum mechanics. The need of having a classical environment in order both to perform the measure and to define the quantum eigenstates, indeed, leads to a great theoretical loophole: Is the classical world necessary to the quantum mechanics or the quantum evolution is the fundamental law?
Besides, the unavailability of the theoretical connection between the quantum and the classical mechanics, that would explain how the laws of physics pass from the deterministic quantum behavior to the classical one (even irreversible), leaves open many questions about how concepts of the classical experience such as, measure, principle of causality, locality, physical state of the external reality, wave and particle behaviors, can be compatible or related to the quantum mechanics.
The connection between the quantum state and the statistical (classical) process of measure is defined by a postulate that, is a matter of fact, makes the quantum mechanics a semi-empirical theory without a self-consistent theoretical framework.
On the other hand, if the wave function is something physically real, then, there must exist a defined mechanism (e.g., the so-called wave function collapse, out of the canonical law of quantum mechanics) expressing the interaction with the observer embedded into a classical universe.
In this case, there also exists the problem about how the Schrödinger equation can be generalized [13, 14] or derived in the frame of such a more general quantum theory [15].
In order to fill this theoretical lack, there exist various interpretations of quantum mechanics like the many-worlds interpretation [17], the Bohmian mechanics [18, 19], the modal interpretation [20], the relational interpretation [21], the consistent histories [22], the transactional interpretation [23, 24], the QBism [25], the Madelung quantum hydrodynamics [26-28] and the decoherence approach [29] .
The Madelung approach (that is a particular case of the Bohmian mechanics [30]) owns the important peculiarities to be both mathematically equivalent to the Schrödinger one [31] and to treat the wave function evolution $\psi = |\psi| e^{i\frac{S}{\hbar}}$ in the classical-like representation as the motion of the mass density $|\psi|^2$ owing the impulse $p_i = \frac{\partial S}{\partial q_i}$. In this way it introduces the concept of trajectories of motion and naturally hosts the notion of physical reality before the measure.
The Madelung description has the advantage to disembogues into the classical mechanics as soon as $\hbar$ and the so-called quantum pseudo-potential are set to zero.
Nevertheless, if we wipe out (by hand) the quantum potential from the quantum hydrodynamic equations in order to obtain the classical mechanics, we also cancel the stationary quantum eigenstates where the total force exerted by the Hamiltonian potential and the quantum one (on the mass density distribution $|\psi|^2$ ) is null. Doing that, we change the nature of the equation of motion. Thence, a more correct and analytic mechanism is needed to pass from the quantum non-local description to the classical one in the frame of the hydrodynamic approach.

Others characteristics of the quantum to classical transition are captured by the decoherence approach that investigates the possibility of obtaining the classical state through the lost of quantum coherence generated by the presence of the environment. The decoherence is shown to be produced into the system by treating it as a sub-part of the overall system, comprehending the environment whose interaction is semi-empirically defined by non unitary interaction [29]. However, this approach is not able to explain how, by having a quantum overall system, the observer can perform the irreversible processes of the statistical measure (and to be quantum de-coupled with the measured system). To overcome this problem, the relational quantum mechanics introduces the super-observer that is not entangled with the overall system [21]. Actually, this "ad hoc" postulate, is unsatisfactory and brings logical contradictions.

From the experimental and numerical simulation point of view, there exist the important evidence that the decoherence and the localization of quantum states come from the interaction with the stochastic fluctuations of the environment [32-35] and/or dark matter.

In this work the authors generalize the Madelung quantum hydrodynamic approach to its stochastic version, where the noise, due to the quantum-mechanical properties of a fluctuating vacuum (in term of curvature associated to dark matter), owns a non-white spectrum showing the emergence of the intrinsic De Broglie physical length into the vacuum.

The work also shows that the stochastic Langevin-Schrodinger equation is derived from the theory for systems whose physical length is much smaller than the De Broglie length.

In the final section the authors analyze how the classical mechanics can be achieved, under appropriate conditions, on a large scale description. The uncertainty principle in the measure process is investigated in the frame of the stochastic quantum hydrodynamic model (SQHM). The paper analyzes how the measure in a classical large-scale system can satisfy both the uncertainty principle and the finite velocity of transmission of light and information.

## 2. The quantum hydrodynamic equation in presence of vacuum dark mass density fluctuations

In the present work we go beyond the flat static solution $R_{\mu\nu} - \frac{1}{2} R g_{\mu\nu} = \frac{8\pi G}{c^4} T_{\nu\mu} = 0$ as for classical matter vacuum, and assume that there is still energy and momentum within the space-time due to the possible presence of gravitational waves that can give a contribution $\frac{8\pi G}{c^4} T'_{\nu\mu} \neq 0$ . Solution to such equations has been introduced by de Sitter that illustrates that matter may not be the only source of gravity and thus wrinkles in the space-time may not be due to matter only.

By considering the vacuum as a fluctuating background, we define the stochastic generalization of the quantum-hydrodynamic equations [26-28,31] that for the wave function $\psi = |\psi| e^{-\frac{iS}{\hbar}}$ are given by the conservation equation for the mass density $n = |\psi|^2$

$$\frac{\partial}{\partial t} n + \frac{\partial}{\partial q_i} ( n \dot{q}_i ) = 0 . \qquad (2.1)$$

and by the motion equations

$$\dot{q}_i = \frac{p_i}{m} = \frac{1}{m}\frac{\partial S_{(q,t)}}{\partial q_i}, \qquad (2.2)$$

$$\dot{p}_i = -\frac{\partial (H + V_{qu_{cl}})}{\partial q_i}, \qquad (2.3)$$

where $H$ is the classical Hamiltonian of the system and where

$$V_{qu} = -\frac{\hbar^2}{2m}\frac{1}{n^{1/2}}\frac{\partial^2 n^{1/2}}{\partial q_i \partial q_i}, \qquad (2.4)$$

The ripples of the vacuum curvature are assumed to manifest themselves by an additional fluctuating mass density distribution (MDD) $¢n_{vac}$

$$n_{tot} \equiv \overline{n} + ¢n_{vac} \qquad (2.5)$$

where $lim_{¢n_{vac} \to 0}\, \overline{n} = n$, that, through the quantum potential

$$V_{qu(n_{tot})} = -\frac{\hbar^2}{2m} n_{tot}^{-1/2} \frac{\partial^2 n_{tot}^{1/2}}{\partial q_i \partial q_i}, \qquad (2.6)$$

leads to the fluctuating force

$$-\frac{\partial V_{qu(n_{tot})}}{\partial q_i}. \qquad (2.7)$$

Being the mass density $¢n_{vac}$ defined positive, the vacuum fluctuations (as a mean $<¢n_{vac}>$) give rise to an additional mass that, owning just the gravitational interaction, is dark matter.

For the purpose of this work, we assume that the vacuum dark matter (DM) does not interact with the physical system (the gravity interaction is disregarded for its weak constant and it is not included in $H$). As far as concerning the dark matter evolution, it is defined by additional (gravitational) motion equation descending by the cosmological dynamics. Nevertheless, we disregard the DM cosmological evolution and assume, for our laboratory macroscopic systems, that the dark mass $<¢n_{vac}>$ is locally uniformly distributed with a constant amplitude of fluctuations $¢n_{(q,t)}$ such as

$$¢n_{vac} \cong <¢n_{vac}> + ¢n_{(q,t)} \qquad (2.8)$$

## 2.1 Spectrum and correlation function of the dark-matter-induced mass density fluctuations

In deriving the characteristics of the quantum potential fluctuations (and hence of its force), we use the condition that the vacuum dark matter, described by the wave function $\Psi_{vac}$, does not interact with the physical system (this due to the weak gravity constant). In this case the wave function $\Psi_{tot}$ of the overall, system reads

$$\Psi_{tot} \cong \Psi \Psi_{vac} \tag{2.1.1}$$

leading to the overall quantum potential

$$\begin{aligned} V_{qu(n_{tot})} &= -\frac{\hbar^2}{2m} |\Psi|^{-1} |\Psi_{vac}|^{-1} \frac{\partial^2 |\Psi| |\Psi_{vac}|}{\partial q_i \partial q_i} = \\ &= -\frac{\hbar^2}{2m}\left( |\Psi|^{-1} \frac{\partial^2 |\Psi|}{\partial q_i \partial q_i} + |\Psi_{vac}|^{-1} \frac{\partial^2 |\Psi_{vac}|}{\partial q_i \partial q_i} + |\Psi|^{-1} |\Psi_{vac}|^{-1} \frac{\partial |\Psi_{vac}|}{\partial q_i} \frac{\partial |\Psi|}{\partial q_i} \right) \end{aligned} \tag{2.1.2}$$

Moreover, given the energy fluctuations

$$\Delta \overline{E}_{qu} = \int_V n_{tot(q,t)} \Delta V_{qu(q,t)} dq \,, \tag{2.1.3}$$

due to the vacuum dark mass density noise of wave-length $\lambda$

$$\Delta n_{vac(\lambda)} = |\Psi_{vac(\lambda)}|^2 \propto cos^2 \frac{2\pi}{\lambda} q \tag{2.1.4}$$

(associated to the dark matter wave-function fluctuation

$$\Psi_{vac} \propto \pm cos \frac{2\pi}{\lambda} q \;) \tag{2.1.5}$$

where

$$\begin{aligned} \Delta V_{qu(q,t)} &= -\frac{\hbar^2}{2m}\left( |\Psi_{vac}|^{-1} \frac{\partial^2 |\Psi_{vac}|}{\partial q_i \partial q_i} + |\Psi|^{-1} |\Psi_{vac}|^{-1} \frac{\partial |\Psi_{vac}|}{\partial q_i} \frac{\partial |\Psi|}{\partial q_i} \right) \\ &= \frac{\hbar^2}{2m}\left( \left(\frac{2\pi}{\lambda}\right)^2 + |\Psi|^{-1} \frac{\partial |\Psi|}{\partial q_i} \left( \pm cos \frac{2\pi}{\lambda} q \right)^{-1} \left( \pm sin \frac{2\pi}{\lambda} q \right) \right), \\ &= \frac{\hbar^2}{2m}\left( \left(\frac{2\pi}{\lambda}\right)^2 + |\Psi|^{-1} \frac{\partial |\Psi|}{\partial q_i} tan \frac{2\pi}{\lambda} q \right) \end{aligned} \tag{2.1.6}$$

at small wave length ($\lambda \ll V^{1/3}$), reads

$$\begin{aligned} \Delta \overline{E}_{qu(\lambda)} &= \frac{1}{\overline{n}_{tot} V} \frac{\hbar^2}{2m} \int_V n_{tot(q,t)} \left( \left(\frac{2\pi}{\lambda}\right)^2 + |\Psi|^{-1} \frac{\partial |\Psi|}{\partial q_i} tan \frac{2\pi}{\lambda} q \right) dq \\ &= \frac{1}{\overline{n}_{tot} V} \frac{\hbar^2}{2m} \left( \left(\frac{2\pi}{\lambda}\right)^2 \int_V n_{tot(q,t)} dq + \int_V n_{tot(q,t)} \left( |\Psi|^{-1} \frac{\partial |\Psi|}{\partial q_i} tan \frac{2\pi}{\lambda} q \right) dq \right) \cong \frac{\hbar^2}{2m} \left(\frac{2\pi}{\lambda}\right)^2 \end{aligned} \tag{2.1.7}$$

(D.12)

where it has been used the normalization condition $\int_V n_{tot(q,t)} dq = \overline{n}_{tot} V$ and where, on large volume, it has been used the approximation

$$lim_{a \to 0} \int_{-\infty}^{\infty} n_{tot(q,t)} \left( |\cdot|^{-1} \frac{\partial |\cdot|}{\partial q_i} tan \frac{2®}{a} q \right) dq \ll \bar{n}_{tot} V \left( \frac{2®}{a} \right)^2 . \qquad (2.1.8)$$

The result (2.1.7) shows that the energy, due to the mass density fluctuations, increases as the inverse squared of $a$ . Being so, the associated quantum potential fluctuations, on very short distance (i.e., $a \to 0$), can lead to unlimited large energy fluctuations even in the case of vanishing noise amplitude (i.e., $T \to 0$).

In order to warrant the convergence of equations (2.2-3, 2.6) to the deterministic limit (2.2-4) of quantum mechanics for $T \to 0$, this behavior imposes the need of a supplemental condition on the spatial correlation function of the noise (we name it $G(a)$).

The derivation of conditions on the noise correlation function shape $G(a)$, brings a quite heavy stochastic calculations [36]. A more simple and straight derivation of $G(a)$ can be obtained by considering the spectrum of the fluctuations.

Since each component of spatial frequency $k = \frac{2®}{a}$ brings the quantum potential energy contribution (2.1.6), the probability of happening

$$p = \boldsymbol{exp}\left[ -\frac{E}{kT} \right], \qquad (2.1.9)$$

by (2.1.7) reads

$$p_{(a)} \propto exp\left[ -\frac{¢\bar{E}_{qu}}{kT} \right]$$

$$= exp\left[ -\frac{\frac{\hbar^2}{2m}\left(\frac{2®}{a}\right)^2}{kT} \right] = exp\left[ -\left(\frac{®a_c}{a}\right)^2 \right] \qquad (2.1.10)$$

where

$$a_c = 2\frac{\hbar}{(2mkT)^{1/2}} \qquad (2.1.11)$$

is the De Broglie length.

From (2.1.10) it comes out that the spatial frequency spectrum $S(k)$

$$S(k) \propto p(\frac{2®}{a}) = exp\left[ -\left(\frac{®a_c}{a}\right)^2 \right] = exp\left[ -\left(\frac{ka_c}{2}\right)^2 \right] \qquad (2.1.12)$$

is not white and the components with wave-length $a$ smaller than $a_c$ go quickly to zero.

Thence, given the mass density noise correlation function, that for the sufficiently general case, to be of practical interest, can be assumed Gaussian with null correlation time, isotropic into the space and independent among different co-ordinates such as

$$< ¢n_{(q_{\sim},t)}, ¢n_{(q+a,t+²)} > = < ¢n_{(q_{\sim})}, ¢n_{(q)} >_{(T)} G(a)¢(²)¢_{\sim} , \qquad (2.1.13)$$

the spatial shape $G(a)$ reads

$$G_{(a)} \propto \int_{-\infty}^{+\infty} exp[ik a] S_{(k)} dk \propto \int_{-\infty}^{+\infty} exp[ik a]\, exp\left[ -\left( k \frac{a_c}{2} \right)^2 \right] dk$$

$$\propto \frac{®^{1/2}}{a_c} exp\left[ -\left( \frac{a}{a_c} \right)^2 \right] \quad . \qquad (2.1.14)$$

The expression (2.1.14) shows that uncorrelated MDD fluctuations on shorter and shorter distance are progressively suppressed by the quantum potential allowing the deterministic quantum mechanics to realize itself for systems whose physical length is much smaller than the De Broglie one.

## 2.2 The stochastic potential approach

The characteristics of the stochastic force noise induced by the fluctuations of the quantum potential (due to the vacuum mass density fluctuations) can be derived by assume the quantum potential as composed by a regular part $< V_{qu(n)} >$ (to be defined) plus the zero mean fluctuating part $V_{st}$ such as

$$V_{qu(n_{tot})} = -\frac{\hbar^2}{2m} n_{tot}^{-1/2} \frac{\partial^2 n_{tot}^{1/2}}{\partial q\, \partial q} = < V_{qu} > + V_{st} \,. \qquad (2.2.1)$$

Moreover, given the force noise

$$m´_{(q,t,T)} = -\frac{\partial V_{st}}{\partial q_i}, \qquad (2.2.2)$$

it is possible to show (see appendix A) that it owns the correlation function

$$< ´_{(q_{\sim},t)} ´_{(q\ +a,t+²)} > = < ´_{(q_{\sim})} ´_{(q\ )} >_{(T)} F(a)¢(²)¢_{\sim} \qquad (2.2.3)$$

with the condition

$$\lim_{T \to 0} < ´_{(q_{\sim})} ´_{(q\ )} >_{(T)} = 0 \qquad (2.2.4)$$

where the spatial shape $F(a)$ is connected to $G(a)$ of the dark matter and where $T$ is the fluctuation amplitude parameter (of DM). Thence, the motion equation acquires the form

$$\dot{q}_i = \frac{p}{m}, \qquad (2.2.5)$$

$$\dot{p}_i = -\frac{\partial (V_{(q)} + < V_{qu(n_{tot})} >)}{\partial q_i} + m´_{(q,t,T)}, \qquad (2.2.6)$$

## 2.3 Correlation function of the quantum force fluctuations for De Broglie-length small-scale systems

As shown in Appendix A, the correlation function of the quantum potential fluctuations, at the smallest order in $\frac{a}{a_c}$, reads

$$< {}^{\prime}{}_{(q\sim,t)}, {}^{\prime}{}_{(q\ +^{a},t+^{2})} > \cong \frac{\hbar^4}{4m^4}\frac{a^2}{a_c^{\ 2}} < ¢\, n_{(q\sim,t)}, ¢\, n_{(q\ +^{a},t+^{2})} >$$

$$= \frac{\hbar^4}{4m^4}\frac{a^2}{a_c^{\ 2}} < ¢\, n_{(q\sim)}, ¢\, n_{(q\sim)} > G(^{a})¢(^{2}) \quad (2.3.1)$$

$$= \frac{\hbar^4}{4m^4}\frac{a^2}{a_c^{\ 2}}\frac{< ¢\, n_{(q\sim)}, ¢\, n_{(q\ )} >_{(T)}}{a_c} exp\left[-\left(\frac{a}{a_c}\right)^2\right] ¢(^{2})¢_{\sim}$$

where $\frac{a}{4®}$ is the boson-boson s-wave scattering length for Lennard-Jones interacting particles (see (3.6) in section 3).

By using the variance (2.3.1), for systems, whose physical length $\mathcal{L}$ is much smaller than the De Broglie length (i.e., $\frac{\mathcal{L}}{a_c} \to 0$), it follows that

$$< {}^{\prime}{}_{(q\sim,t)}, {}^{\prime}{}_{(q\ +^{a},t+^{2})} >_{(T)} = lim_{\ a_c \to\infty} < \ddot{x}_{(q\sim,t)}, \ddot{x}_{(q\ +^{a},t+^{2})} >$$

$$= \frac{\hbar^4}{4m^4}\frac{a^2}{a_c^{\ 2}} lim_{a_c\to\infty} \frac{< ¢\, n_{(q\sim)}, ¢\, n_{(q\ )} >_{(T)}}{a_c} exp\left[-\left(\frac{a}{a_c}\right)^2\right] ¢(^{2})¢_{\sim} \quad (2.3.2)$$

$$\cong \frac{\hbar^4}{4m^4}\frac{a^2}{a_c^{\ 2}}\frac{< ¢\, n_{(q\sim)}, ¢\, n_{(q\ )} >}{a_c} ¢(^{2})¢_{\sim} = \tilde{D}¢(^{2})¢_{\sim}$$

where

$$\tilde{D} = \frac{\hbar^4}{4m^4}\frac{a^2}{a_c^{\ 2}}\frac{< ¢\, n_{(q\sim)}, ¢\, n_{(q\sim)} >}{a_c} \quad (2.3.3.)$$

where $\left[< ¢\, n_{(q\sim)}, ¢\, n_{(q\sim)} >\right] = l^{-5}t = m^{-5}s$

Besides, about the regular regular part $< V_{qu} >$, for microscopic systems, without loss of generality, we can pose

$$< V_{qu} > = -\frac{\hbar^2}{2m} < n_{tot}^{\ -1/2} \frac{\partial^2 n_{tot}^{\ 1/2}}{\partial q\ \partial q} > = -\frac{\hbar^2}{2m}\frac{1}{n^{1/2}}\frac{\partial^2 n^{1/2}}{\partial q\ \partial q} + \overline{V}_{st} = V_{qu(n)} + \overline{V}_{st} \quad (2.3.4)$$

where the spatial probability mass density (PMD) $n_{(q,t)}$ reads

$$n_{(q,t)} = \int \mathcal{N}(q,p,t)d^3p \quad (2.3.5)$$

where

$$\mathcal{N}(q,\dot{q},t) = \int P(q,\dot{q},z,\dot{z},|t,0)\mathcal{N}_{(z,\dot{z},0)} d^3z d^3\dot{z} \qquad (2.3.6)$$

(where $P(q,\dot{q},z,\dot{z},|t,0)$ is the PTF of the phase space Smolukowski equation (see (C.3) in Appendix C) for the Brownian process

$$\ddot{q}_{j(t)} = -\copyright \dot{q}_{j(t)} - \frac{1}{m}\frac{\partial\left(V_{(q)} + V_{qu(n)}\right)}{\partial q_j} + \copyright D^{1/2}\P_{(t)}, \qquad (2.3.7)$$

where, as shown in Appendix B, at first order of approximation in $\dot{q}_{(t)}$, the term $\bar{V}_{st}$ generates the

$$-\frac{1}{m}\frac{\partial \bar{V}_{st}}{\partial q_j} \cong -\copyright \dot{q}_{(t)} \qquad (2.3.8)$$

and where $D = \frac{\tilde{D}}{\copyright^2}$ .

Moreover, the relation between the friction coefficient $= m\copyright_{(\frac{1}{D})}$ and the diffusion coefficient $D$ , at the first order in the series expansion

$$\copyright_{\left(\frac{1}{D}\right)} \cong 0 + \copyright'_{(0)}\frac{1}{D}, \qquad (2.3.9)$$

can generally read

$$\copyright = \sim \frac{2kT}{mD} + \left(O_{(1/D)}\right)^2 \qquad (2.3.10)$$

where $\sim$ is a numerical parameter that measures how the quantum hydrodynamic trajectories of motion are perturbed by fluctuations (leading to quantum decoherence and to energy dissipation). This parameter is specific for each considered system since the sensibility of the system to fluctuations is related to the Lyapunov exponents of their classical trajectories of motion. This aspect goes beyond the purpose of this work and (2.3.10) is semi-empirically assumed here.

For $\sim = 0$ we have a system that maintains the quantum coherence with no dissipation (e.g., as happens in the deterministic limit).

For $\sim = 1$ we retrieve the Einstein relation $D = \frac{2kT}{m\copyright}$ that holds for the so called "dust matter" (the mass density, constituted by monomolecular dust, representing the classical MDD limit of the Madelung quantum hydrodynamics [31]).

In the general case of a system submitted to fluctuations, neither in quantum deterministic limit nor in the classical one, we assume that

$$0 < \sim \le 1.$$

In order the SQHM theory comprehends the canonical quantum mechanics in the deterministic limit (i.e., $\frac{a_c}{\mathcal{L}} \to \infty$, where $\mathcal{L}$ is the physical length of the system), it must hold both that

$$lim_{T\to 0} D = 0 \qquad (2.3.11)$$

and that

$$lim_{T\to 0}\text{©} = 0 \,. \qquad (2.3.12)$$

Given that , by (2.3.2)

$$\tilde{D} = <\text{´}_{(q\sim,t)},\text{´}_{(q\sim)}> \; = \frac{\hbar^4}{4m^4}\frac{a^2}{\text{a}_c^{\,2}}\frac{<\text{¢}n_{(q\sim)},\text{¢}n_{(q\sim)}>}{\text{a}_c} \qquad (2.3.13)$$

we obtain

$$D = \frac{\left(2\sim kT\right)^2}{<\text{¢}n_{(q\sim)},\text{¢}n_{(q\sim)}>_{(T)}}\frac{4m^2\text{a}_c^{\,3}}{a^2\hbar^4} = 2^6\frac{\sim^2}{a^2<\text{¢}n_{(q\sim)},\text{¢}n_{(q\;\;)}>_{(T)}}\frac{\left(mkT\right)^{1/2}}{\hbar} \qquad (2.3.14)$$

$$\text{©} = 2^{-5}\frac{\hbar}{m^{3/2}}\frac{a^2<\text{¢}n_{(q\sim)},\text{¢}n_{(q\sim)}>_{(T)}\left(kT\right)^{1/2}}{\sim} \quad , \qquad (2.3.15)$$

That, by posing

$$lim_{T\to 0}<\text{¢}n_{(q\sim)},\text{¢}n_{(q\sim)}>_{(T)} = 2^5 n_0^2\left(kT\right)^{\text{i}} \;\; (\text{with } \text{i} \ge 0) \qquad (2.3.16)$$

leads to

$$lim_{T\to 0}D = 2\frac{\sim^2}{n_0^2 a^2}\frac{m^{1/2}}{\hbar}\left(kT\right)^{1/2-\text{i}} \qquad (2.3.17)$$

$$lim_{T\to 0}\text{©} = \frac{\hbar}{m^{3/2}}\frac{n_0^2}{\sim}\left(kT\right)^{\text{i}+\frac{1}{2}} \quad . \qquad (2.3.18)$$

Moreover, by by assuming in the limit of small fluctuations amplitude

$$lim_{T\to 0}\sim \;\approx\; \sim_0\left(kT\right)^{\text{¥}} \,, \quad \text{¥} > 0 \,, \qquad (2.3.19)$$

it follows that

$$D = 2\frac{\sim_0^{\,2}}{n_0^2}\frac{m^{1/2}}{\hbar}\left(kT\right)^{1/2-\text{i}+2\text{¥}} \qquad (2.3.20)$$

$$\text{©} = \frac{\hbar}{m^{3/2}}\frac{n_0^2}{\sim_0}\left(kT\right)^{\text{i}+\frac{1}{2}-\text{¥}} \quad , \qquad (2.3.21)$$

so that (2.3.11-12) are satisfied by

$$\frac{\text{i}-\frac{1}{2}}{2} < \text{¥} < \text{i}+\frac{1}{2} \quad . \qquad (2.3.22)$$

For instance, for $\text{i} = 1/2$ and $\text{¥} = 1/2$ it follows that

$$D = 2\frac{\sim_0^{\;2}}{n_0^2}\frac{m^{1/2}}{\hbar}kT \qquad (2.3.23)$$

$$© = \frac{\hbar}{m^{3/2}}\frac{n_0^2}{\sim_0}\left(kT\right)^{1/2} \quad ,. \qquad (2.3.24)$$

$$\tilde{D} = ©^2 D = 2\frac{\hbar}{m^{5/2}}n_0^{\;2}\left(kT\right)^2 ; \qquad (2.3.25)$$

For $¡ = 0$ and $¥ = 1/4$, we obtain

$$D = 2\frac{\sim_0^{\;2}}{n_0^2}\frac{m^{1/2}}{\hbar}kT \qquad (2.3.26)$$

$$© = \frac{\hbar}{m^{3/2}}\frac{n_0^2}{\sim_0}\left(kT\right)^{1/4} \quad , \qquad (2.3.27)$$

$$\tilde{D} = ©^2 D = 2\frac{\hbar}{m^{5/2}}n_0^{\;2}\left(kT\right)^{3/2} \qquad (2.3.28)$$

Furthermore, by posing

$$D = ¥_D \frac{\hbar}{2m}\left(\frac{£}{a_c}\right)^p \qquad (2.3.29)$$

and

$$© = \sim \frac{4kT}{¥_D \hbar}\left(\frac{a_c}{£}\right)^p \qquad (2.3.30)$$

where $¥_D$ is the dimensionless constant

$$¥_D = \frac{4}{£^p}\frac{\sim_0^{\;2}}{n_0^2}\frac{m^{\frac{3-p}{2}}}{\hbar^{2-p}}\left(kT\right)^{1/2-¡+2¥-p/2}, \qquad (2.3.31)$$

that for $p = 2$ reads

$$¥_D = 4\frac{m^{1/2}}{£^2}\frac{\sim_0^{\;2}}{n_0^2}\left(kT\right)^{2¥-¡-1/2} \qquad (2.3.32)$$

and for $2¥ = ¡ + 1/2$ (2.3.23-4, 2.23.6-7)) gives

$$¥_D = 4\frac{m^{1/2}}{£^2}\frac{\sim_0^{\;2}}{n_0^2}. \qquad (2.3.33)$$

**2.4 The generalized Langevin-Schrödinger equation for De Broglie-length systems**

The quantum-hydrodynamic equation (2.3.7) for the complex field

$$\cdot_{(q,t)} = |\cdot_{(q,t)}|\, exp\left[S_{(q,t)}\right] = n_{(q,t)}^{\;1/2}\, exp\left[S_{(q,t)}\right] \qquad (2.4.1)$$

where

$$\frac{1}{m}\frac{\partial S}{\partial q_{\sim}} = \dot{q} \qquad (2.4.2)$$

reads

$$\ddot{q}_{\sim} = \frac{1}{m}\frac{d}{dt}\frac{\partial S}{\partial q_{\sim}} = \frac{\partial}{\partial t}\frac{\partial S}{\partial q_{\sim}} + \frac{1}{m}\frac{\partial^2 S}{\partial q_{\sim}\partial q}\frac{\partial S}{\partial q} = \frac{\partial}{\partial q_{\sim}}\left(\frac{\partial S}{\partial t} + \frac{1}{m}\frac{\partial S}{\partial q}\frac{\partial S}{\partial q}\right)$$

$$= -\frac{1}{m}\frac{\partial\left(V_{(q)} - \frac{\hbar^2}{2m|\cdot|^{1/2}}\frac{\partial^2 |\cdot|^{1/2}}{\partial q\,\partial q} + \quad S - q^{1/2} q^{1/2} m\tilde{D}^{1/2} ¶_{\sim(t)}\right)}{\partial q_{\sim}} \qquad (2.4.3)$$

leading to the partial stochastic differential equation

$$m\frac{\partial S}{\partial t} + \frac{\partial S}{\partial q}\frac{\partial S}{\partial q} = -\left(V_{(q)} - \frac{\hbar^2}{2m|\cdot|^{1/2}}\frac{\partial^2 |\cdot|^{1/2}}{\partial q\,\partial q} + \quad S - q^{1/2} q^{1/2} m\tilde{D}^{1/2} ¶_{\sim(t)} + C_{(t)}\right) \qquad (2.4.4)$$

(C.2.4)

### 2.4.1. Introducing the environment

The presence of unnoticeable small dark matter fluctuations, even negligible on the ordinary scale systems, is suffice to lead to a finite De Broglie length that is much smaller than the cosmological scale allowing the quantum decoherence and the emergence of the classical behavior into the universe.

The possibility of dividing the universe in classical sub-parts, allows to correctly introduce the existence of the environment. Besides, given that the action of force noise of the environment on the MDD derivatives generates an increase of the energy of the quantum potential in the same way as the dark matter, it follows that the spatial shape of the correlation function of the force noise of the environment owns the same form of that one of the dark matter.

Thence, in presence of the physical environment it is possible to assume the stochastic interaction

$$V_{ext_{st}} = \quad {}_{ext}S - q^{1/2} q^{1/2} \tilde{D}_{ext}{}^{1/2} ¶_{(t)} \qquad (2.4.1.1)$$

from which it follows that

$$V_{(q)} + V_{ext_{st}} + \quad S - q^{1/2} q^{1/2} m\tilde{D}^{1/2} ¶_{(t)} + C_{(t)} =$$
$$V_{(q)} + \bar{\ } S - q^{1/2} q^{1/2} m\bar{D}^{1/2} ¶_{(t)} + C_{(t)} \qquad (2.4.1.2)$$

where, by assuming both $\tilde{D} \ll \tilde{D}_{ext}$ .and $\quad \ll \quad _{ext}$ , it follows that

$$\bar{D} = \tilde{D} + \tilde{D}_{ext} \cong \tilde{D}_{ext} \qquad (2.4.1.3)$$

and

$$\bar{\ } \cong \quad + \quad _{ext} \cong \quad _{ext}\, , \qquad (2.4.1.4)$$

and that all preceding formulas can be retained with the substitution

$$\tilde{D} \to \tilde{D}_{ext} \qquad (2.4.1.5)$$

$$\quad \to \quad _{ext} \qquad (2.4.1.6)$$

It must be noted that, in absence of dark matter fluctuations, the quantum-decoupling of the environment cannot be assumed and that $V_{ext_{st}}$ cannot be formulated in the form of (2.4.1.1) . Thence, (2.4.1.5-6) can be assumed only in a physical vacuum that, following the general relativity, is constituted by a fluctuating geometrical background.

**2.5 The Langevin-Schrödinger equation**

Once the dark matter makes possible to have the classical environment, a system of microscopic physical length $\mathcal{L}$ (i.e., $\frac{\mathcal{L}}{a_c} \to 0$ ) $n$ obeys to the conservation equation (see (C.3.9, C.3.11) in appendix C)

$$\lim_{\frac{\mathcal{L}}{a_c}\to 0}\left( \partial_t n_{(q,t)} + \frac{\partial( n_{(q,t)} < \dot{q} >)}{\partial q_i} + Q_{diss_{(q,t)}} \right)$$
$$\cong \lim_{\frac{\mathcal{L}}{a_c}\to 0}\left( \partial_t n_{(q,t)} + \frac{\partial( n_{(q,t)}\dot{q})}{\partial q_i} + Q_{diss_{(q,t)}} \right) = 0 \qquad (2.4.1.7)$$

where (see ( C.3.11) in appendix C)

$$Q_{diss_{(q,t)}} = \int \begin{pmatrix} 0 \\ \frac{1}{2}\frac{\partial \tilde{D}\mathcal{N}_{(q,p,t)}}{\partial p_{\sim}} + \ldots + \frac{1}{n!}\sum_{h=2}^{\infty} \frac{\partial^k C^{(k)}_{\sim i \ldots\ldots \pounds}\, \mathcal{N}_{(q,p,t)}}{\underbrace{\partial p_{i} \ldots \partial p_{\pounds}}_{(k-terms)}} \end{pmatrix} d^{3h} p \qquad (2.4.1.8)$$

and where it has been used the identity $\lim_{\frac{a_c}{\mathcal{L}}\to\infty} < \dot{q} >= \dot{q}$ . Thence, by (2.4.1.7) it follows that

$$\frac{\partial |\cdot|}{\partial t} = -\frac{1}{m}\frac{\partial |\cdot|}{\partial q_{\sim}}\frac{\partial S}{\partial q_{\sim}} - \frac{1}{2m}|\cdot| \frac{\partial}{\partial q_{\sim}}\frac{\partial S}{\partial q_{\sim}} + \frac{Q_{diss_{(q,t)}}}{|\cdot|} \qquad (2.4.1.9)$$

leading to the generalized Langevin-Schrodinger equation (LSE)

$$-i\hbar\frac{\partial}{\partial t}\cdot = \frac{\hbar^2}{2m}\frac{\partial^2}{\partial q\,\partial q}\cdot - \left( V_{(q)} + S - q^{1/2} q^{1/2}\tilde{D}^{1/2}\P_{(t)} + i\frac{Q_{diss_{(q,t)}}}{|\cdot|^2} + C_{(t)} \right)\cdot \qquad (2.4.1.10)$$

(C.2.7)

## 2.4.2. Robust quantum systems and the canonical LSE

Moreover, since for a quantum system that owns $\sim_0 \to 0$ and hence © $\gg \tilde{D}$, able to strongly maintain its quantum coherence (we name it "robust" quantum systems), it holds

$$\lim_{\sim_0\to 0} Q_{diss_{(q,t)}} \to 0 \ , \qquad (2.4.2.1)$$

and, for

$$T<\propto \frac{\sim_0^{\frac{2}{2¡+1-2¥}}}{k} \qquad (2.4.2.2)$$

(e.g., $T<\frac{\sim_0^{2}}{k}$ for $¡=1/2$ and $¥=1/2$), also that

$$lim_{\sim_0 \to 0} © = finite \,, \qquad (2.4.2.3)$$

for sufficiently small temperature, the term $i\frac{Q_{diss(q,t)}}{|\cdot|^2}$ can be disregarded in (2.4.1.10) and the LSE reads

$$lim_{\sim_0, T\to 0}\, i\hbar\frac{\partial}{\partial t}\cdot = -\frac{\hbar^2}{2m}\frac{\partial^2}{\partial q\ \partial q}\cdot + \left(V_{(q)} + \quad S - q^{1/2} q^{1/2} \tilde{D}^{1/2} ¶_{(t)} + C_{(t)}\right)\cdot \quad .(2.4.2.4)\ (LSE)$$

As already observed, the sensibility of the system to fluctuations is related to the Lyapunov exponents of its classical trajectories of motion. This aspect goes beyond the purpose of this work and is not analyzed here. It is suffice to say that for linear systems (non classically chaotic) $\sim_0 \to 0$ and the (LSE) can be applied to them.

Generally speaking, for the case of classically chaotic systems, the LSE (2.4.1.10) has to be considered.

Moreover, it must be noted that the SLE description is made possible by the integrability of the velocity field $\dot{q} = \frac{1}{m}\frac{\partial S}{\partial q}$ that can be warranted in small scale (slightly perturbed) quantum system but it may fall in macroscopic large-scale system whose velocity field can be non-integrable.

## 2.5 The SQHM with adiabatic elimination of fast variables

For slow kinetics with the characteristic time $^2{}_{ch}$ satisfying the condition

$$^2{}_{ch} \gg \frac{1}{©} = \frac{m^{3/2}\sim_0}{\hbar n_0{}^2\sqrt{kT}} = \frac{¥_D}{16\sim}\frac{£^2 m}{\hbar}\,, \qquad (2.5.1)$$

equation (2.3.7) (for $T>0$ ) reduces to

$$\dot{q}_{(t)} = -\frac{¥_D\hbar}{4\sim mkT}\left(\frac{£}{a_c}\right)^{p}\frac{\partial\left(V_{(q)}+V_{qu}\right)}{\partial q} + ¥_D^{1/2}\left(\frac{£}{a_c}\right)^{p/2}\left(\frac{\hbar}{2m}\right)^{1/2} ¶_{(t)}. \quad (2.5.2)$$

that for $p=2$ reads

$$\dot{q}_{(t)} = -\frac{\text{¥}_D m \text{£}^2}{16\hbar \text{~}}\frac{\partial\left(V_{(q)}+V_{qu}\right)}{\partial q} + \text{¥}_D^{1/2}\frac{\text{£}}{2}\sqrt{\frac{kT}{\hbar}}\text{¶}_{(t)}$$

$$= -\frac{m^{3/2}\text{~}_0}{\hbar {n_0}^2\sqrt{kT}}\frac{\partial\left(V_{(q)}+V_{qu}\right)}{\partial q} + m^{1/4}\hbar\frac{\text{~}_0}{n_0}\sqrt{\frac{2kT}{m}}\text{¶}_{(t)} \qquad (2.5.3)$$

For a quantum system with $m \approx 10^{-30} Kg$ and $\text{£} \approx 10^{-10} m$ , equation (2.5.2) can be applied to kinetics with characteristic time down to

$$2_{ch} \gg \frac{\text{¥}_D m \text{£}^2}{16\hbar \text{~}} \approx \frac{\text{¥}_D}{2\text{~}_{(T)}}10^{-17} s \qquad (2.5.4)$$

that being

$$\text{¥}_D = 4\frac{m^{1/2}}{\text{£}^2}\frac{\text{~}_0^2}{n_0^2} \approx 4\times 10^5\frac{\text{~}_0^2}{n_0^2} \qquad (2.5.5)$$

gives

$$2_{ch} \gg \frac{\text{~}_0}{n_0^2}\frac{m^{3/2}}{4\hbar\left(kT\right)^{\text{¥}}} \approx \frac{2\text{~}_0}{n_0^2\left(kT\right)^{\text{¥}}}10^{-12} s \qquad (2.5.6)$$

It is worth mentioning that equation (2.5.3) leads to a simplified Smolukowski equation that is only a function of the space variables [37].

# 3. Emerging of the classical behavior on coarse-grained large scale

Is matter of fact that, if the quantum potential is canceled by hand in the quantum hydrodynamic equations of motion (2.1-3), the classical mechanical equation of motion emerges [28]. Even if this is true, this operation is not mathematically correct since it changes the characteristics of the QHA equations. Doing so, the stationary configurations (i.e., eigenstates) are wiped out because we cancel the balancing of the quantum potential force against the Hamiltonian force [37] that generates the stationary condition. Thence, an even small quantum potential cannot be neglected into the deterministic QHA model.

On the contrary, in the SQHM it is possible to correctly neglect the quantum potential (at least in classically chaotic systems) when its force is much smaller than the noise ´ such as

$$. \left|\frac{1}{m}\frac{\partial V_{qu(n)}}{\partial q_i}\right| \ll \left|\text{´}_{(q,t,T)}\right| . \qquad (3.1)$$

When the non-local force generated by the quantum potential is quite small (respect to the fluctuations amplitude) so that

$$\left|\frac{1}{m}\frac{\partial V_{qu(n)}}{\partial q_i}\right| \ll \text{©}\left(\frac{\text{£}}{a_c}\right)\left(\text{¥}_D\frac{\hbar}{2m}\right)^{1/2} = \text{©}\left(\frac{\text{£}\sqrt{mkT}}{2\hbar}\right)\left(\text{¥}_D\frac{\hbar}{2m}\right)^{1/2} , \qquad (3.2)$$

its effect can be disregarded in (2.3.7) .

Besides, even if the noise $\text{´}_{(q,t,T)}$ has zero mean, the mean of the quantum potential fluctuations $\bar{V}_{st(n,S)} \cong$ $S$ is not zero, and the stochastic sequence of inputs of noise alters the coherent reconstruction of the quantum superposition of state by the dissipative force $-\text{©}\dot{q}_{(t)}$ in (2-3-7). Moreover, by observing that

$$©\left(\frac{£}{ª_c}\right)\left(¥_D\frac{\hbar}{2m}\right)^{1/2}¶_{(t)} \tag{3.3}$$

grows with the scale of the system (i,e., $\frac{£}{ª_c} \to \infty$ for macroscopic systems), condition (3.2) is satisfied if

$$lim_{\frac{q}{ª_c}\to\infty} |\frac{1}{m}\frac{\partial V_{qu(n_{(q)})}}{\partial q_i}| = limited \tag{3.4}$$

and the classical behavior can emerge in systems of sufficiently large physical length. Actually, in order to have a large-scale description, completely free from quantum effects, we can more strictly require

$$lim_{\frac{q}{ª_c}\to\infty} |\frac{1}{m}\frac{\partial V_{qu(n_{(q)})}}{\partial q_i}| = \frac{1}{m}\sqrt{\frac{\partial V_{qu(n_{(q)})}}{\partial q_i}\frac{\partial V_{qu(n_{(q)})}}{\partial q_i}} = 0\,. \tag{3.5}$$

By observing that for linear systems

$$lim_{q\to\infty} V_{qu(q)} \propto q^2\,, \tag{3.6}$$

from (105) the SQHM shows that they do never have a macroscopic classical phase. Generally speaking, stronger the Hamiltonian potential higher the wave function localization and larger the quantum potential behavior at infinity [38]. Given the MDD

$$|\cdot|^2 \propto exp\left[-P^k{}_{(q)}\right] \tag{3.7}$$

where $P^k{}_{(q)}$ is a polynomial of order *k*, in order to have a finite quantum potential range of interaction it must be $k < \frac{3}{2}$ (it results $k = 2$ for uni-dimensional linear interaction). Actually, since the linear interaction is not maintained up to infinity (for energetic reason, a finite bound energy requires a weaker than linear interaction such as $lim_{q\to\infty} V_{(q)} \to 0$ ), there exists a large-scale classical description when the physical length of the system is much larger than the range of linear interaction. A physical example comes from solids owning a quantum lattice. If we look at the intermolecular features where the interaction is linear, the behavior is quantum (such as the x-ray diffraction shows), but if we look at their macroscopic properties (e.g., low-frequency acoustic wave propagation) the classical behavior is shown.

For instance, systems that interact by the Lennard-Jones potential for which the long distance wave function reads [39]

$$lim_{r\to\infty} |\cdot| \propto a^{-1/2}\frac{1}{r} \tag{3.8}$$

that leads to the quantum potential

$$lim_{r\to\infty} V_{qu(n)} = lim_{q\to\infty}\frac{\hbar^2}{2m}\frac{1}{|\cdot|}\frac{\partial^2|\cdot|}{\partial r\partial r} = \frac{1}{r^2} = \frac{\hbar^2}{m}a|\cdot|^2 \tag{3.9}$$

and to the quantum force

$$lim_{r\to\infty}\frac{\partial V_{qu(n)}}{\partial r}=lim_{q\to\infty}\frac{\hbar^2}{2m}\frac{\partial}{\partial r}\frac{1}{|\cdot|}\frac{\partial^2|\cdot|}{\partial r\partial r}=\frac{\hbar^2}{2m}\frac{\partial}{\partial r}r\frac{\partial^2\frac{1}{r}}{\partial r\partial r}=-2\frac{\hbar^2}{m}\frac{1}{r^3}, \quad (3.10)$$

the large scale classical behavior can appear [38, 40] in a sufficiently rarefied phase (see section 4.4). It is interesting to note that in (3.6) the quantum potential acquires the form of the hard sphere potential of the pseudo potential Hamiltonian model of the Gross-Pitaevskii equation [15, 41] where $\frac{a}{4®}$ is the boson-boson s-wave scattering length.

By observing that, in order to fulfill the condition (3.5) a sufficient condition reads

$$\int_0^\infty r^{-1}|\frac{1}{m}\frac{\partial V_{qu(n(q))}}{\partial q_i}|_{(r,¯,¨)}\, dr = limited \qquad \forall ¯,¨ \qquad , \quad (3.11)$$

it is possible to define the quantum potential range of interaction [38]

$$a_{qu}=a_c\frac{\int_0^\infty r^{-1}|\frac{\partial V_{qu(n(q))}}{\partial q_i}|_{(r,¯,¨)}\, dr}{|\frac{\partial V_{qu(n(q))}}{\partial q_i}|_{(r=a_c,¯,¨)}} \quad (3.12)$$

that gives a measure of the physical length of the quantum non-local effects.

The convergence of the integral (3.11) for $r\to 0$ is warranted for L-J type potentials since, near the equilibrium point ($r=0$), the L-J interaction is linear and being $lim_{r\to 0}V_{qu(n(q\to 0))}\propto r^2$ it follows that $lim_{r\to 0}r^{-1}|\frac{\partial V_{qu(n(q))}}{\partial q_i}|_{(r,¯,¨)}\simeq constant$ . (3.13)

## 3.1 From micro to macro description

By discretizing the phase space conservation equation given by the current equation (2.2.5-6, 2.4, 2.3.4, 2.3.8) for the system of N coupled particles [42], it is possible to obtain the quantum hydrodynamic master equation for a macroscopic system of a huge number of molecules.

In order to obtain the macroscopic description, we may procede by discretization of the stochastic quantum hydrodynamic equations

$$\frac{\partial n}{\partial t}+\frac{\partial(n\dot{q})}{\partial q_i}=0$$

$$\dot{q}_i=\frac{p}{m}, (20)$$

$$\dot{p}_i=-m©\dot{q}_{j(t)}-\frac{\partial\left(V_{(q)}+V_{qu(n)}\right)}{\partial q_j}+m´_{(q,t,T)}, \quad (D.5) \quad (3.1.1)$$

For the purpose of this work, we can exemplify the to the simpler case kinetics with a characteristic time $2_{ch}$ larger than

$$
{}^{2}{}_{ch} \gg \frac{1}{©} = \frac{\hbar^{2} \sim_{0}}{«_{0}^{2}\sqrt{kT}}, \qquad (20)\ (3.1.2)
$$

leading (for $T > 0$ ) to

$$
\frac{\partial n}{\partial t} + \frac{\partial (n\dot{q})}{\partial q_i} = 0
$$

$$
\dot{q}_{j(t)} = -\frac{1}{m©}\frac{\partial\left(V_{(q)} + V_{qu(n)}\right)}{\partial q_j} + \frac{1}{©}\,{}'_{(q,t,T)}
$$

Given the conserved equation in the local $j$-th cell of side $l$ with the current $J_{j(q,t)} = n\dot{q}_{j(t)}$ that reads

$$
\begin{aligned}
J_{(q,t)} = n\dot{q}_{(t)} &= -n\left( \frac{m\mathcal{L}^{2}}{4\sim}\partial_{q}\left(V_{(q)} + V_{qu}\right) + {}'_{(q,t,T)} \right) \\
&\cong -n\left( \frac{m\mathcal{L}^{2}}{4\sim}\partial_{q}\left(V_{(q)} + V_{qu}\right) + \Phi_{(q,t)}¶_{(t)} \right)
\end{aligned}, \qquad (3.1.3)
$$

by posing

$$
x_j = l^{3} n_{tot(q_j,t)}, \qquad (3.1.4)
$$

the discrete spatial generalization of the SDE (3.1.3)reads

$$
dx_j = -\sum_{m,k} \mathcal{D}'_{jm} \frac{m\mathcal{L}^{2}}{4\sim_{j}} x_m \mathcal{D}_{mk}\left(V_k + V_{qu_k}\right)dt + \sum_{k} \mathcal{D}''_{jk} x_k \Phi_k dW_{k(t)} \qquad (3.1.5)
$$

where $V_k = V_{(q_k)}$, $V_{qu_k} = V_{(n_{(q_k)})qu}$, $\Phi_k = \Phi_{(q_k,t)}$ and where the terms $\mathcal{D}_{jk}$, $\mathcal{D}'_{jk}$ and $\mathcal{D}''_{jk}$ are matrices of coefficients that corresponding to the discrete approximation of the derivaties $\partial_q$ and where

$$
lim_{l \to 0}\, l^{-6} < \Phi_j, \Phi_k > = < {}'_{(q_j)}, {}'_{(q_j)} >_{(T)} G_{(l(k-j))}. \qquad (3.1.6)
$$

Generally speaking, the form of the overall interaction (classicl plus quantum) $V_{(q)_k} + V_{qu_k}$ steming by the $k$-th cell, depends by the physical system and its evolution.

For instance, by assuming $l = \mathcal{L} \gg ª_c, ª_{qu}$ (where $\mathcal{L}^3$ is the available volume per molecule), for a system of sufficiently rarefied phase of L-J interacting particles with a asymptotically vanishing quantum potential (3.9-10) , the quantum potential interaction between adjacent cells is null and, hence, the classical master equation is obtained.

Here, generally speaking we observe that, given the range of interaction of the quantum potential $ª_{qu}$, the De Broglie length $ª_c$, and the system size $\mathcal{L}$ ($\mathcal{L}^3 \sim$ the mean available volume per molecule), we can generally distinguish the cases:

i. $\mathcal{L} \gg ª_{qu}, ª_c$

ii. $\mathcal{L} > ª_{qu}, ª_c$

iii. $ª_{qu} > \mathcal{L} > ª_c$

iv. $a_{c} > \mathcal{L}$

Typically, for the L-J potential the quantum potential range of interaction $a_{qu}$ extends itself a little bit further than the linear zone around the equilibrium position $r_0$ (let' say up to $r_0 + \Delta$ ).
By using this approximation for the L-J interaction, so that for $r < r_0 + \Delta$

$$\frac{\partial V_{qu(n)}}{\partial q} \approx \sim r \, , \tag{3.1.7}$$

and for $r > r_0 + \Delta$ [39]

$$-\frac{\partial V_{qu(n)}}{\partial q} \approx 2\frac{\hbar^2}{m}\frac{1}{r^3} \, , \tag{3.1.8}$$

$a_{qu}$ reads

$$a_{qu} \approx a_{c} \left( \frac{\int_0^{r_0+\Delta} dr}{a_{c}} + \frac{\int_{r_0+\Delta}^{\infty} \frac{1}{r^4} dr}{\frac{1}{a_{c}^{3}}} \right) = r_0 + a + \frac{a_{c}^{4}}{3\left(r_0 + \Delta\right)^3} \tag{3.1.9}$$

that, since for $T \gg 4°k$ and microscopic mass $m \approx 10^{-30} Kg$ , $a_{c} \ll r_0 \approx 10^{-9m}$, we obtain

$$a_{qu} \approx r_0 + \Delta \tag{3.1.10}$$

Thence, the rarefied phases owing $\mathcal{L} \gg a_{qu} \simeq r_0 + \Delta \gg a_{c}$, for particles interacting by a Lennard-Jones potential, is fully classic since the mean molecular distance $\mathcal{L}$ is much larger both than the De Broglie length and the quantum potential length of interaction $a_{qu}$.

The second case $\mathcal{L} > a_{qu} \simeq r_0 + \Delta \gg a_{c}$ refers to dense phases (e.g., fluid phase) that still own a classical behavior since, as a mean, the particle are distant each-other more than the range of interaction of the quantum potential. The inter-particle distance mostly lies in the non-linear range of L-J interaction [38].

The case "iii" $r_0 + \Delta \simeq a_{qu} > \mathcal{L} \gg a_{c}$ applies when the neighboring molecules lie in the linear range of the intermolecular potential at a distance smaller than the non-local quantum potential interaction $a_{qu}$.

The observables on such physical scale show quantum behavior (e.g., the Bragg's diffraction of the atomic lattice).

In the case "iv" $a_{c} > \mathcal{L}$, when the condensed fluid phase (i.e., $\mathcal{L} > r_0 + \Delta$ ) persists down to a very low temperature so that the De Broglie length becomes larger than the mean intermolecular distance ($T \sim< 4°k$ for typical intermolecular distance of order of $10^{-9} m$), the fluid shows an extreme decrease of molecular viscosity [38]. The super-fluidity is induced by the quantum potential interaction between the molecules [40]

By changing the temperature and, accordingly, both $a_{c}$ and the mean inter-molecular distance $\mathcal{L}$ , we can have quantum-to-classic phase transition in the case iii and iv, respectively:

i. $\frac{\mathcal{L}}{a_{qu}} \rightarrow 1+£$ (with $a_c < a_{qu}$) solid-fluid transition with melting of crystalline lattice (e.g., ice -water transition [38])

ii. $\frac{\mathcal{L}}{a_c} \rightarrow 1+£$ (with $a_{qu} < a_c$) superfluid-fluid transition (e.g., $He_4$ lambda point [38, 40])

### 3.2. Measurement process and quantum decoherence

The SQHA model shows that the measure is not necessarily a decoherent process by itself: The sensing part of the measuring apparatus (the pointer) and the measured system may have a canonical quantum interaction that, after the measurement when the measuring apparatus is brought to the infinity (at a distance much beyond $a_c$), ends. Then the reading and the treatment of the "pointer" state is done by the measurement apparatus: This process is practically a classic irreversible process (with a defined arrow of time) leading to the macroscopic output of the measure.
On the other hand, the decoherence is necessary for the measurement process in order to have, both before the initial time and after the final one, quantum-decoupling between the measurement apparatus and the system in order to collect a statistical ensemble of data from repeated measures.

### 3.3. Minimum measurements uncertainty

If for physical length much smaller than $a_c$ any system approaches the quantum deterministic behavior and behave as a wave so that its sub-parts are not independent each-other, it follows that in order to perform the measurement (with independence between the measuring apparatus and the measured system) it is necessary that they are far apart (at least) more than $a_c$ and hence, for the finite speed of propagation of interactions and information, the measure process must last longer than the time

$$2 = \frac{a_c}{c} = \frac{2\hbar}{(2mc^2kT)^{1/2}} \quad . \qquad (3.3.1)$$

For $a_{qu} > a_c$ the measurement time can be even bigger than (3.3.1) but not less.
Moreover, since higher the amplitude of the noise $T$ lower the value of $a_c$ and higher the fluctuations of the energy measurements $\Delta E_{(T)}$, it follows that the minimum duration of the measurement $2 = \frac{a_c}{c}$ multiplied by the precision of the energy measurement $\Delta E_{(T)}$ has a lower bond.

Given the Gaussian property of the noise (2.3.2), we have that the mean value of the energy fluctuation is $\Delta E_{(T)} = \frac{1}{2}kT$. Thence, for the non-relativistic case ($mc^2 >> kT$) a particle of mass *m* owns an energy variance $\Delta E$

$$\Delta E \approx (<(mc^2 + \Delta E_{(T)})^2 - (mc^2)^2>)^{1/2} \cong (2mc^2\Delta E_{(T)})^{1/2} \cong (mc^2kT)^{1/2} \quad (3.3.2)$$

from which it follows that

$$\Delta E\Delta t > \Delta E\Delta 2 = \frac{(mc^2kT)^{1/2}\,\mathrm{a}_c}{c} = \sqrt{2}\,\frac{h}{®}, \tag{3.3.3}$$

It is worth noting that the product $\Delta E\Delta 2$ is constant since the growing of the energy variance with the square root of $T$ is exactly compensated the decrease of the minimum time $2$ of measurement

The same result is achieved if we derive the experimental uncertainty between the position and momentum of a particle of mass *m* in the quantum fluctuating hydrodynamic model.
If we measure both the spatial position of a particle with a precision $\Delta L > \mathrm{a}_c$ (so that we are able to not perturb the quantum configuration of the measured system) and the variance $\Delta p$ of the modulus of its relativistic momentum $(p^{«}p_{«})^{1/2} = mc$ due to the fluctuations that reads

$$\Delta p \approx (<(mc+\frac{\Delta E_{(T)}}{c})^2-(mc)^2>)^{1/2} \cong (<(mc)^2+2m\Delta E-(mc)^2>)^{1/2}$$
$$\cong (2m\Delta E_{(T)})^{1/2} \cong (mkT)^{1/2} \tag{3.3.4}$$

we obtain the experimental uncertainty

$$\Delta L\Delta p > \mathrm{a}_c(mkT)^{1/2} = \frac{\sqrt{2}}{®}h \tag{3.3.5}$$

If we measure the spatial position with a precision $\Delta L < \mathrm{a}_c$, we have to perturb the quantum state. Due to the increase of the spatial confinement of the wave function (by increasing the environmental temperature or by an external potential), the increase of both the quantum potential energy and its fluctuations are generated so that the final particle momentum gets a variance $\Delta p$ higher than (3.3.5).
It is worth mentioning that the SQHM leads to the *minimum measurements uncertainty* as a consequence of the relativistic postulate of finite speed of light and information.
Even if the quantum deterministic behavior ($\mathrm{a}_c \to \infty$) in the low velocity limit ($c \to \infty$) leads to the undetermined inequalities

$$2 \geq \frac{\mathrm{a}_c}{c} \tag{3.3.6}$$

$$\Delta E \cong (mc^2kT)^{1/2} = \sqrt{2}\hbar\frac{c}{\mathrm{a}_c} \tag{3.3.7}$$

their product

$$2\Delta E \geq 2\hbar \tag{3.3.8}$$

remains defined and constitutes the *minimum uncertainty* of the quantum deterministic limit. Beside, (3.3.6) in the relativistic limit shows that the duration of the measurement process in the deterministic limit becomes infinite. Being it endless, it is not possible to perform it in the canonical quantum mechanical universe.

Moreover, since non-locality is confined in domains of physical length smaller than $\mathrm{a}_c$ and information cannot be transferred faster than the light speed (otherwise also the uncertainty principle will be violated) the local realism is obtained in the coarse-grained large scale physics and the paradox of a "spooky action at a distance [43]" is limited on a distance of order of $\mathrm{a}_c$ or of $\mathrm{a}_{qu}$.

The above result holds for particles with rest mass different from zero, while for determining the length of non local interaction (entanglement) of the photon ( $a_c \to \infty$ ?) the relativistic generalization of the SQHM is required.

### 3.4. Field of Application of the SQHM

The theory that describes how the quantum entanglement is maintained up to a certain distance and how it can be maximized, can lead to important improvements in the development of materials for high-temperature superconductors and Q bits systems.
Moreover, the theory owing a self-defined quantum correlation distance can be also very important in defining different regimes of chemical kinetics in complex reactions and phase transitions.
Besides, the SQHM can furnish an analytical self-consistent theoretical model for mesoscale phenomena and quantum irreversibility.

## 4. Conclusions

The SQHM describes how the quantum dynamics realizes itself in a vacuum whose metric fluctuates. In this scenario the canonical quantum mechanics is the limiting description achieved in a flat static vacuum.
Figuratively, in a 3-dimensional space time, where the space can be represented by the surface of a see with very small ripples (instead by a flat static plane) the non local interaction of quantum mechanics breaks down on large scale and, in huge systems of weakly bounded particles, the classical mechanics emerges.
The SQHM, shows that in the physical fluctuating vacuum, the spatial spectrum of the noise is not white and it owns the De Broglie characteristic length. Due to this fact, the quantum entanglement is effective in systems whose physical length is much smaller than such a length. The model shows that the non-local quantum interactions may extend themselves up to a finite distance in the case of non-linear weakly bonded systems.
The SQHM shows that the Schroedinger equation can be derived by taking into account, at the first order of approximation, the effect of fluctuations on microscopic systems.
The derivation of the LSE from the general SQHM allows to define its basic assumptions and its range of applicability limited to microscopic systems whose physical length is much smaller than the De Broglie one.
The SQHM shows that the minimum uncertainty condition is satisfied during the process of measurement in a fluctuating environment and that it can have a finite duration.
The theory shows that the minimum uncertainty in the measurement process is satisfied if, and only if, interactions and information do not travel faster than the speed of light, making compatible the relativistic postulate (at the base of the large scale locality) with the non-local quantum interactions at the micro-scale.
The SQHM makes compatible the hydrodynamic description of quantum mechanics with the decoherence approach showing that the quantum potential is not able to maintain the quantum coherence in presence of fluctuations, generating a frictional force leading to a relaxation process (decoherence). The theory shows that the superposition of states does not physically exist in macroscopic systems made up of molecules and atoms interacting by long-range weak potentials such as the Lennard-Jones one.

# References

1. T. Young, Phil. Trans. R. Soc. Lond. 94, 1 (1804).
2. R. P. Feynmann, The Feynman Lectures on Physics, Volume 3, (AddisonWesley, 1963).
3. G. Auletta, Foundations and Interpretation of Quantum Mechanics, (World Scientific, 2001).
4. G. Greenstein and A. G. Zajonc, The Quantum Challenge, (Jones and Bartlett Publishers, Boston, 2005), 2nd ed.
5. P. Shadbolt, J. C. F. Mathews, A. Laing and J. L. OBrien, Nature Physics 10, 278 (2014).
6. C. Josson, Am. J. Phys, 42, 4 (1974).
7. A. Zeilinger, R. Gahler, C.G. Shull, W. Treimer and W. Mampe, Rev. Mod. Phys. 60, 1067 (1988).
8. O. Carnal and J. Mlynek, Phys. Rev. Lett. 66, 2689 (1991).
9. W. Sch¨ollkopf and J.P. Toennies, Science 266, 1345 (1994).
10. M. Arndt, O. Nairz, J. Vos-Andreae, C. Keller, G. van der Zouw and A. Zeilinger, Nature 401, 680 (1999).
11. O. Nairz, M. Arndt and A. Zeilinger, Am. J. Phys. 71, 319 (2003).
12. M. Born, The statistical interpretation of quantum mechanics - Nobel Lecture, December 11, 1954.
13. G.C. Ghirardi, A. Rimini and T. Weber, Phys. Rev. D. 34, 470 (1986).
14. G.C. Ghirardi, Collapse Theories, Edward N. Zalta (ed.), The Stanford Encyclopedia of Philosophy (Fall 2018 Edition).
15. P.P.Pitaevskii,"Vortex lines in an Imperfect Bose Gas", Soviet Physics JETP. 1961;13(2):451–454.
16. H. Everette, Rev. Mod. Phys. 29, 454 (1957).
17. L. Vaidman, Many-Worlds Interpretation of Quantum Mechanics, Edward N. Zalta (ed.), The Stanford Encyclopedia of Philosophy (Fall 2018 Edition).
18. D. Bohm, Phys. Rev. 85, 166 (1952).
19. S. Goldstein, Bohmian Mechanics, Edward N. Zalta (ed.), The Stanford Encyclopedia of Philosophy (Summer 2017 Edition).
20. O. Lombardi and D. Dieks, Modal Interpretations of Quantum Mechanics, Edward N. Zalta (ed.), The Stanford Encyclopedia of Philosophy (Spring 2017 Edition).
21. F. Laudisa and C. Rovelli, Relational Quantum Mechanics, Edward N. Zalta (ed.), The Stanford Encyclopedia of Philosophy (Summer 2013 Edition).
22. R.B. Griffiths, Consistent Quantum Theory, Cambridge University Press (2003).
23. J.G. Cramer, Phys. Rev. D 22, 362 (1980).
24. J.G. Cramer, The Quantum Handshake: Entanglement, Non-locality and Transaction, Springer Verlag (2016).
25. H. C. von Baeyer, QBism: The Future of Quantum Physics, Cambridge, Harvard University Press, (2016)
26. Madelung, E.:. Z. Phys. 40, 322-6 (1926).
27. Jánossy, L.: Zum hydrodynamischen Modell der Quantenmechanik. Z. Phys. 169, 79 (1962).
28. Weiner, J.H., *Statistical Mechanics of Elasticity* (John Wiley & Sons, New York, 1983), p. 315-7.
29. Lidar, D. A.; Chuang, I. L.; Whaley, K. B. (1998). "Decoherence-Free Subspaces for Quantum Computation". Physical Review Letters. **81** (12): 2594–2597.
30. Tsekov, R., Bohmian mechanics versus Madelung quantum hydrodynamics, arXiv:0904.0723v8 [quantum-phys] (2011).
31. I. Bialyniki-Birula, M., Cieplak, J., Kaminski, "Theory of Quanta", Oxford University press, Ny, (1992) 87-111.
32. Cerruti, N.R., Lakshminarayan, A., Lefebvre, T.H., Tomsovic, S.: Exploring phase space localization of chaotic eigenstates via parametric variation. Phys. Rev. E 63, 016208 (2000).
33. E. Calzetta and B. L. Hu, Quantum Fluctuations, Decoherence of the Mean Field, and Structure Formation in the Early Universe, Phys.Rev. D, **52**, 6770-6788, (1995).

34. C., Wang, P., Bonifacio, R., Bingham, J., T., Mendonca, Detection of quantum decoherence due to spacetime fluctuations, 37th COSPAR Scientific Assembly. Held 13-20 July 2008, in Montréal, Canada., p.3390.
35. F., C., Lombardo , P. I. Villar, Decoherence induced by zero-point fluctuations in quantum Brownian motion, Physics Letters A 336 (2005) 16–24.
36. P., Chiarelli, Can fluctuating quantum states acquire the classical behavior on large scale? J. Adv. Phys. 2013; **2**, 139-163.
37. P., Chiarelli, Stability of quantum eigenstates and kinetics of wave function collapse in a fluctuating vacuum , in progress.
38. P., Chiarelli, Quantum to Classical Transition in the Stochastic Hydrodynamic Analogy: The Explanation of the Lindemann Relation and the Analogies Between the Maximum of Density at He Lambda Point and that One at Water-Ice Phase Transition, Physical Review & Research International, 3(4): 348-366, 2013.
39. Bressanini D. "An accurate and compact wave function for the 4He dimer", EPL. 2011;96.
40. P., Chiarelli, The quantum potential: the missing interaction in the density maximum of $He^4$ at the lambda point?, Am. J. Phys. Chem.. **2**(6) (2014) 122-131.
41. Gross EP. Structure of a quantized vortex in boson systems. Il Nuovo Cimento,1961;20(3):454–457. doi:10.1007/BF02731494.
42. C.W. Gardiner, *Handbook of Stochastic Method*, 2 nd Edition, Springer, (1985) ISBN 3-540-61634.9, pp. 331-41 .
43. A. Einstein, B. Podolsky and N. Rosen, Can Quantum-Mechanical Description of Physical Reality Be Considered Complete? Phys, Rev., 47, 777-80 (1935).
44. Weiner, J. H. and Forman, R.: Rate theory for solids. V. Quantum Brownian-motion model. Phys. Rev. B 10, 325 (1974).20.
45. Ruggiero P. and Zannetti, M., Critical Phenomena at T=0 and Stochastic Quantization Phys. Rev. Lett. 47, 1231 (1981);
46. Ruggiero P. and Zannetti, M., Microscopic derivation of the stochastic process for the quantum Brownian oscillator, Phys. Rev. A 28, 987 (1983).
47. Y.B., Rumer, M.S., Ryvkin Thermodynamics, Statistical Physics, and Kinetics. Moscow: Mir Publishers; 1980; 269.

# Appendix A

In orderto derive the correlation function of the quantum potential fluctuations, we assume that the the dark matter density (MD) fluctuations $\text{¢}n_{(q,t)}$ own an amplitude that is very much smaller than the MD of the physical system $n$ and hence it follows that

$$
\begin{aligned}
V_{qu} &= -\frac{\hbar^2}{2m}\frac{1}{n_{tot}^{1/2}}\frac{\partial^2 n_{tot}^{1/2}}{\partial q_{\text{¥}}\partial q_{\text{¥}}} = -\frac{\hbar^2}{2m}\frac{1}{\left(\bar{n}+\text{¢}n\right)^{1/2}}\frac{\partial^2\left(\bar{n}+\text{¢}n\right)^{1/2}}{\partial q_{\text{¥}}\partial q_{\text{¥}}} \\
&\cong -\frac{\hbar^2}{2m}\frac{1}{\bar{n}^{1/2}\left(1+\frac{\text{¢}n}{2\bar{n}}\right)}\frac{\partial^2\bar{n}^{1/2}\left(1+\frac{\text{¢}n}{2\bar{n}}\right)}{\partial q_{\text{¥}}\partial q_{\text{¥}}} \\
&= -\frac{\hbar^2}{2m}\frac{1}{\bar{n}^{1/2}\left(1+\frac{\text{¢}n}{2\bar{n}}\right)}\left(\frac{\left(1+\frac{\text{¢}n}{2\bar{n}}\right)\partial^2\bar{n}^{1/2}}{\partial q_{\text{¥}}\partial q_{\text{¥}}}+2\frac{\partial}{\partial q_{\text{¥}}}\left(1+\frac{\text{¢}n}{2\bar{n}}\right)\frac{\partial\bar{n}^{1/2}}{\partial q_{\text{¥}}} + \bar{n}^{1/2}\frac{\partial^2\left(1+\frac{\text{¢}n}{2\bar{n}}\right)}{\partial q_{\text{¥}}\partial q_{\text{¥}}}\right) \quad \text{(A.1)} \\
&= -\frac{\hbar^2}{2m}\left(\frac{\partial^2\bar{n}^{1/2}}{\partial q_{\text{¥}}\partial q_{\text{¥}}}+\frac{\partial\frac{\text{¢}n}{2\bar{n}}}{\partial q_{\text{¥}}}\frac{\partial ln\bar{n}}{\partial q_{\text{¥}}}+\frac{1}{\left(1+\frac{\text{¢}n}{2\bar{n}}\right)}\frac{\partial^2\frac{\text{¢}n}{2\bar{n}}}{\partial q_{\text{¥}}\partial q_{\text{¥}}}\right)
\end{aligned}
$$

where

$$lim_{T\to 0}\,\bar{n} = n\,. \qquad \text{(A.2)}$$

Therefore,

$$
\begin{aligned}
\frac{4m^2}{\hbar^4} &< V_{qu}, V_{qu} > = < \frac{\partial^2\bar{n}^{1/2}}{\partial q_{\text{¥}}\partial q_{\text{¥}}}, \frac{\partial^2\bar{n}^{1/2}}{\partial q_{\text{¥}}\partial q_{\text{¥}}} > +2 < \frac{\partial^2\bar{n}^{1/2}}{\partial q_{\text{¥}}\partial q_{\text{¥}}}, \frac{\partial\frac{\text{¢}n}{2\bar{n}}}{\partial q_{\text{¥}}}\frac{\partial ln\bar{n}}{\partial q_{\text{¥}}} > \\
&+2 < \frac{\partial^2\bar{n}^{1/2}}{\partial q_{\text{¥}}\partial q_{\text{¥}}}, \left(1-\frac{\text{¢}n}{2\bar{n}}\right)\frac{\partial^2\frac{\text{¢}n}{2\bar{n}}}{\partial q_{\text{¥}}\partial q_{\text{¥}}} > + < \frac{\partial\frac{\text{¢}n}{2\bar{n}}}{\partial q_{\text{¥}}}\frac{\partial ln\bar{n}}{\partial q_{\text{¥}}}, \frac{\partial\frac{\text{¢}n}{2\bar{n}}}{\partial q_{\text{¥}}}\frac{\partial ln\bar{n}}{\partial q_{\text{¥}}} > \qquad \text{(A.3)} \\
&+2 < \frac{\partial\frac{\text{¢}n}{2\bar{n}}}{\partial q_{\text{¥}}}\frac{\partial ln\bar{n}}{\partial q_{\text{¥}}}, \left(1-\frac{\text{¢}n}{2\bar{n}}\right)\frac{\partial^2\frac{\text{¢}n}{2\bar{n}}}{\partial q_{\text{¥}}\partial q_{\text{¥}}} > + < \left(1-\frac{\text{¢}n}{2\bar{n}}\right)\frac{\partial^2\frac{\text{¢}n}{2\bar{n}}}{\partial q_{\text{¥}}\partial q_{\text{¥}}}, \left(1-\frac{\text{¢}n}{2\bar{n}}\right)\frac{\partial^2\frac{\text{¢}n}{2\bar{n}}}{\partial q_{\text{¥}}\partial q_{\text{¥}}} >
\end{aligned}
$$

that since the mean value $\bar{n}$ is not random, leads to

$$
lim_{\frac{a}{c}\to 0}\frac{4m^2}{\hbar^4}<V_{qu(q)},V_{qu(q'=q+a)}>
$$

$$
=lim_{\frac{a}{c}\to 0}<\frac{\partial\frac{¢n}{2\bar{n}}}{\partial q_{¥}}\frac{\partial ln\bar{n}}{\partial q_{¥}},\frac{\partial'\frac{¢n}{2\bar{n}}}{\partial q'_{¥}}\frac{\partial' ln\bar{n}}{\partial q'_{¥}}>+2<\frac{\partial\frac{¢n}{2\bar{n}}}{\partial q_{¥}}\frac{\partial ln\bar{n}}{\partial q_{¥}},\left(1-\frac{¢n}{2\bar{n}}\right)\frac{\partial'^2\frac{¢n}{2\bar{n}}}{\partial' q_{¥}\partial' q_{¥}}>
$$

$$
+<\left(1-\frac{¢n}{2\bar{n}}\right)\frac{\partial^2\frac{¢n}{2\bar{n}}}{\partial q_{¥}\partial q_{¥}},\left(1-\frac{¢n}{2\bar{n}}\right)\frac{\partial'^2\frac{¢n}{2\bar{n}}}{\partial' q_{¥}\partial' q_{¥}}> \qquad (A.4)
$$

$$
=lim_{\frac{a}{c}\to 0}\frac{\partial ln\bar{n}}{\partial q_{¥}}\frac{\partial' ln\bar{n}}{\partial q'_{¥}}<\frac{\partial\frac{¢n}{2\bar{n}}}{\partial q_{¥}},\frac{\partial'\frac{¢n}{2\bar{n}}}{\partial q'_{¥}}>+2\frac{\partial ln\bar{n}}{\partial q_{¥}}<\frac{\partial\frac{¢n}{2\bar{n}}}{\partial q_{¥}},\left(1-\frac{¢n}{2\bar{n}}\right)\frac{\partial'^2\frac{¢n}{2\bar{n}}}{\partial' q_{¥}\partial' q_{¥}}>
$$

$$
+<\left(1-\frac{¢n}{2\bar{n}}\right)\frac{\partial^2\frac{¢n}{2\bar{n}}}{\partial q_{¥}\partial q_{¥}},\left(1-\frac{¢n}{2\bar{n}}\right)\frac{\partial'^2\frac{¢n}{2\bar{n}}}{\partial' q_{¥}\partial' q_{¥}}>
$$

that at first order in $a$ reads

$$
lim_{\frac{a}{c}\to 0}\frac{4m^2}{\hbar^4}<V_{qu(q)},V_{qu(q'=q+a)}>
$$

$$
\cong lim_{\frac{a}{c}\to 0}\frac{\partial ln\bar{n}}{\partial q_{¥}}\frac{\partial' ln\bar{n}}{\partial q'_{¥}}\frac{\partial}{\partial q_{¥}}\frac{\partial'}{\partial q'_{¥}}<\frac{¢n}{2\bar{n}},\frac{¢n_{(q')}}{2\bar{n}_{(q')}}>+2\frac{\partial ln\bar{n}}{\partial q_{¥}}\frac{\partial}{\partial q_{¥}}\frac{\partial'^2}{\partial' q_{¥}\partial' q_{¥}}<\frac{¢n}{2\bar{n}},\frac{¢n_{(q')}}{2\bar{n}_{(q')}}>
$$

$$
+\frac{\partial^2}{\partial q_{¥}\partial q_{¥}}\frac{\partial'^2}{\partial' q_{¥}\partial' q_{¥}}<\frac{¢n}{2\bar{n}},\frac{¢n_{(q')}}{2\bar{n}_{(q')}}>
$$

$$
\cong lim_{\frac{a}{c}\to 0}\frac{\partial ln\bar{n}}{\partial q_{¥}}\frac{\partial' ln\bar{n}}{\partial q'_{¥}}\frac{1}{4\bar{n}\,\bar{n}_{(q')}}\frac{\partial}{\partial q_{¥}}\frac{\partial'}{\partial q'_{¥}}<¢n,¢n_{(q')}>+\frac{\partial ln\bar{n}}{\partial q_{¥}}\frac{\partial' ln\bar{n}}{\partial q'_{¥}}\frac{\partial\frac{1}{2\bar{n}}}{\partial q_{i}}\frac{\partial'\frac{1}{2\bar{n}_{(q')}}}{\partial q'_{i}}<¢n,¢n_{(q')}>
$$

$$
+2\frac{\partial ln\bar{n}}{\partial q_{¥}}\frac{\partial}{\partial q_{¥}}\frac{\partial'^2}{\partial q'_{i}\,\partial q'_{i}}<\frac{¢n}{2\bar{n}},\frac{¢n_{(q')}}{2\bar{n}_{(q')}}>+\frac{\partial^2}{\partial q_{¥}\partial q_{¥}}\frac{\partial'^2}{\partial q'_{i}\,\partial q'_{i}}<\frac{¢n}{2\bar{n}},\frac{¢n_{(q')}}{2\bar{n}_{(q')}}>
$$

(A.5)

Given that the terms with first derivatives $\frac{\partial}{\partial q_{¥}}$ and $\frac{\partial}{\partial' q_{¥}}$ give terms proportional to $q-q'=a$, in the limit of $a\to 0$ they are null and thence it follows that

$$lim_{\frac{ª}{ª_c}\to 0} \frac{4m^2}{\hbar^4} < V_{qu(q)}, V_{qu(q'=q+ª)} >$$

$$\cong lim_{\frac{ª}{ª_c}\to 0} \frac{1}{8\overline{n}^3}\frac{1}{8\overline{n}^3_{(q')}}\left(\frac{\partial\overline{n}}{\partial q_¥}\frac{\partial'\overline{n}}{\partial q'_¥}\right)^2 < ¢n, ¢n_{(q')} >$$

$$+\frac{1}{2\overline{n}}\frac{1}{2\overline{n}_{(q')}}\frac{\partial^2}{\partial q_¥ \partial q_¥}\frac{\partial'^2}{\partial q'_i\, \partial q'_i} < ¢n, ¢n_{(q')} > \qquad \text{(A.6)}$$

$$\cong lim_{\frac{ª}{ª_c}\to 0} \frac{1}{8\overline{n}^3}\frac{1}{8\overline{n}^3_{(q')}}\left(\frac{\partial\overline{n}}{\partial q_¥}\frac{\partial'\overline{n}}{\partial q'_¥}\right)^2 < ¢n, ¢n_{(q')} >$$

$$+\frac{1}{2\overline{n}}\frac{1}{2\overline{n}_{(q')}}\frac{1}{ª_c^4} < ¢n, ¢n_{(q')} >$$

and, given that $\frac{1}{ª_c^4} \propto (kT)^2$, for very low temperature, it follows that

$$lim_{T\to 0}\, lim_{\frac{ª}{ª_c}\to 0} < V_{qu(q)}, V_{qu(q'=q+ª)} > \cong \frac{1}{2^6}\frac{\hbar^4}{4m^2}\left(\frac{1}{\overline{n}^{3/2}}\frac{\partial\overline{n}}{\partial q_¥}\frac{1}{\overline{n}^{3/2}_{(q')}}\frac{\partial\overline{n}_{(q')}}{\partial q'_¥}\right)^2 < ¢n, ¢n_{(q')} > \qquad \text{(A.7)}$$

As far as it concernes the quantm potetntial force fluctuations, the zero order term can be generally assumed of the form

$$lim_{T\to 0}\, lim_{\frac{ª}{ª_c}\to 0} < \frac{\partial}{\partial q_¥}V_{qu(q)}, \frac{\partial}{\partial q'_¥}V_{qu(q'=q+ª)} > \cong f_{(q,q')} < ¢n, ¢n_{(q')} > \qquad \text{(A.8)}$$

Since $lim_{¢n \ll \overline{n}}\, \overline{n} \to n$, for Lennard-Jones potential we have that

I. $lim_{r/r_0\to\infty}\, \overline{n} \cong lim_{r/r_0\to\infty}\, n \cong lim_{r/r_0\to\infty} |\cdot|^2 \propto a^{-1}\frac{1}{r^2}$ (A.9)

where $\frac{a}{4®}$ is the boson-boson s-wave scattering length. (see (3.6) in section 3), and hence that

II. $$\frac{1}{2^6}\left(\frac{1}{\overline{n}^{3/2}}\frac{\partial\overline{n}}{\partial r}\frac{1}{\overline{n}^{3/2}_{(r')}}\frac{\partial\overline{n}_{(r')}}{\partial r'}\right)^2 \propto a^2\left(r^3 r^{-3} r'^3\, r'^{-3}\right)^2 \propto a^2 \qquad \text{(A.10)}$$

for $r > r_0 + \Delta$.

Moreover, by assuming in the linear range of interaction for $r < r_0 + \Delta$, the Gaussian localization

$$\overline{n} = \overline{n}_0\, exp - \frac{\overline{r}^2}{2\Delta r^2} \qquad \text{(A.11)}$$

where $\overline{r} = r - r_0$, it follows that the diffusion coefficient owns a parabolic behavior

$$\frac{1}{2^6}\left(\frac{\bar{r}}{\Delta r^2}\frac{1}{\bar{n}^{3/2}}\bar{n}\frac{\overline{r'}}{\Delta r'^2}\frac{1}{\bar{n}^{3/2}_{(r')}}\bar{n}_{(r')}\right)^2=\frac{1}{\bar{n}_0^2}\left(\frac{\bar{r}\overline{r'}}{\Delta r^2\Delta r'^2}exp\frac{\bar{r}^2}{4\Delta r^2}+\frac{\overline{r'}^2}{4\Delta r'^2}\right)^2$$

$$\cong\frac{1}{\bar{n}_0^2}\left(\frac{\bar{r}\overline{r'}}{\Delta r^2\Delta r'^2}\left(1+\frac{\bar{r}^2}{4\Delta r^2}+\frac{\overline{r'}^2}{4\Delta r'^2}\right)\right)^2\cong\frac{1}{\bar{n}_0^2}\left(\frac{\bar{r}\overline{r'}}{\Delta r^2\Delta r'^2}\right)^2<\left(\frac{\bar{r}}{\Delta^2}\right)^2\left(\frac{\overline{r'}}{\Delta^2}\right)^2 \tag{A.12}$$

that tends to zero for $\bar{r}\to 0$ or $\overline{r'}\to 0$.

Moreover, given that $\Delta r\sim\Delta$ and that for $r<r_0+\Delta<<\mathcal{L}$ (i.e., $G_{(a)}\to 0$) the ratio $\frac{¢n}{\bar{n}}$ reaches the lowest value (since about all the mass is localized there), the wave function is poorly perturbed by MDD fluctuations (and is well described by the deterministic quantum limit), we can ssume (A.10) over all the space to obtain

$$lim_{T\to 0}\, lim_{\frac{a}{a_c}\to 0} < V_{qu(q)}, V_{qu(q'=q+a)} > \cong a^2\frac{\hbar^4}{4m^2} lim_{T\to 0}\, lim_{\frac{a}{a_c}\to 0} < ¢n_{(q)}, ¢n_{(q')} >$$

$$=\frac{\hbar^4}{4m^4}a^2\frac{< ¢n_{(q)}, ¢n_{(q)} >}{a_c}\; lim_{T\to 0}\, lim_{\frac{a}{a_c}\to 0} exp\left[-\left(\frac{a}{a_c}\right)^2\right]¢(^2\,) \tag{A.13}$$

$$=\frac{\hbar^4}{4m^4}a^2\frac{< ¢n, ¢n >}{a_c}\;¢(^2\,)$$

As far as it concerns the force correlation function, in this case we obtain

$$< ´_{i(q\sim,t)}, ´_{£(q\;\;+a,t+^2)} >= lim_{T\to 0}\, lim_{\frac{a}{a_c}\to 0} < \frac{\partial}{\partial q_i}V_{qu(q\sim)}, \frac{\partial}{\partial q'_£}V_{qu(q'\;\;=q\;\;+a)} >$$

$$= lim_{T\to 0}\, lim_{\frac{a}{a_c}\to 0}\frac{\partial}{\partial q_i}\frac{\partial}{\partial q'_£} < V_{qu(q\sim)}, V_{qu(q'\;\;=q\;\;+a)} >$$

$$= lim_{T\to 0}\, lim_{\frac{a}{a_c}\to 0}\frac{\hbar^4}{4m^4}\frac{a^2}{a_c^2}\frac{< ¢n, ¢n >}{a_c} lim_{T\to 0}\, lim_{\frac{a}{a_c}\to 0} exp\left[-\left(\frac{a}{a_c}\right)^2\right]¢(^2\,)¢_\sim\;¢_{i£} \tag{A.14}$$

$$=\frac{\hbar^4}{4m^4}\frac{a^2}{a_c^2}\frac{< ¢n, ¢n >}{a_c}¢(^2\,)¢_\sim\;¢_{i£}=\tilde{D}¢(^2\,)¢_\sim\;¢_{i£}$$

# Appendix B

## The irreversible force induced by fluctuations in small scale systems

In order to obtain the explicit expression of the term

$$\frac{\partial < V_{qu(n_{tot})} >}{\partial q_{\sim}} \tag{B.1}$$

let's start by equation (2.2.6 )

$$\ddot{q}_{\sim} = -\frac{1}{m} \frac{\partial \left( V_{(q)} + V_{qu(n_{tot})} \right)}{\partial q_{\sim}} \tag{B.2}$$

that we can rearrange as

$$\ddot{q}_{\sim} = -\frac{1}{m} - \frac{\partial \left( V_{(q)} + V_{qu(n)} + \frac{\hbar^2}{2m} \left( \frac{1}{n^{1/2}} \frac{\partial^2 n^{1/2}}{\partial q \, \partial q} - \frac{1}{n_{tot}} \frac{\partial^2 n_{tot}}{\partial q \, \partial q} \right) \right)}{\partial q_{\sim}} \tag{B.3}$$

where the term

$$\frac{\partial}{\partial q_{\sim}} \left( \frac{1}{n^{1/2}} \frac{\partial^2 n^{1/2}}{\partial q \, \partial q} - \frac{1}{n_{tot}} \frac{\partial^2 n_{tot}}{\partial q \, \partial q} \right) \tag{B.4}$$

in (B.3) generates an additional acceleration respect to the deterministic case leading to a change of the velocity field $\dot{q}$ of the mass density. It is noteworthy that, in the deterministic case (B.4) becomes null and $lim_{D \to 0} n_{tot} \equiv n$ (actually, in thre limit of small fluctuations (i.e., small size systems), $n$ is close to the value of the deterministic limit of the eigenstates).

Moreover, by observing that in the stationary states (i.e., $\overline{\dot{q}} = 0$, the analogouses of the eigenstates of the deterministic limit [37] (let's name them quasi-eigenstates), the mean MDD $\overline{n_{tot}}$ does not changes with time and both

$$\overline{n_{tot(q)}} = lim_{\Delta t \to \infty} \frac{1}{\Delta t} \int_{t-\frac{\Delta t}{2}}^{t\frac{\Delta t}{2}} n_{tot(q,ꝛ)} dꝛ = n_{(q)} + < ¢ n_{vac} >= n_{(q)} + const \tag{B.5}$$

and

$$\overline{\dot{q}}_{(q(t),t)} = lim_{\Delta t \to \infty} \frac{1}{\Delta t} \int_{t-\frac{\Delta t}{2}}^{t\frac{\Delta t}{2}} \dot{q} dꝛ = 0, \tag{B.6}$$

approaching the stationary state (i.e., $\overline{\dot{q}} \to 0$ ) it follows that

$$\overline{\frac{\partial}{\partial q_{\sim}} \left( \frac{1}{n^{1/2}} \frac{\partial^2 n^{1/2}}{\partial q \, \partial q} - \frac{1}{n_{tot}} \frac{\partial^2 n_{tot}}{\partial q \, \partial q} \right)} = \left( \frac{\partial}{\partial q_{\sim}} \frac{1}{n^{1/2}} \frac{\partial^2 n^{1/2}}{\partial q \, \partial q} - \overline{\frac{\partial}{\partial q_{\sim}} \frac{1}{n_{tot}} \frac{\partial^2 n_{tot}}{\partial q \, \partial q}} \right) \to 0 \text{,(B.7)}$$

and thence, generally speaking, for small $\overline{\dot{q}}$ , sufficiently close to the stationary quasi-eigenstates, (B.4) can be developed in the series approximation

$$\frac{\partial}{\partial q_{\sim}} \left( \frac{1}{n^{1/2}} \frac{\partial^2 n^{1/2}}{\partial q \, \partial q} - \frac{1}{n_{tot}} \frac{\partial^2 n_{tot}}{\partial q \, \partial q} \right) = A_0 + A_1 \dot{q} + .... + A_n \dot{q}^n \tag{B.8}$$

where $A_0$ is a stochastic noise whose mean $\overline{A_0}$ is defined by the stationary state condition

$$\overline{\frac{\partial}{\partial q_\sim}\left(\frac{1}{n^{1/2}}\frac{\partial^2 n^{1/2}}{\partial q\ \partial q}-\frac{1}{n_{tot}}\frac{\partial^2 n_{tot}}{\partial q\ \partial q}\right)}=\left(\frac{\partial}{\partial q_\sim}\frac{1}{n^{1/2}}\frac{\partial^2 n^{1/2}}{\partial q\ \partial q}-\overline{\frac{\partial}{\partial q_\sim}\frac{1}{n_{tot}}\frac{\partial^2 n_{tot}}{\partial q\ \partial q}}\right)=0 \quad .\text{(B.9)}$$

$$=\overline{A}_{0(q,t)}+\sim_{1(q,t)}\overline{\dot{q}}+....+\sim_{n(q,t)}\overline{\dot{q}^n}=\overline{A}_0=0$$

Thence, at first order in $\dot{q}$, close to the deterministic limit of quantum mechanics (i.e., $\frac{\mathcal{L}}{a_c}\rightarrow 0$),

leads to

$$-\frac{\hbar^2}{2m}\frac{\partial}{\partial q_\sim}\left(\frac{1}{n^{1/2}}\frac{\partial^2 n^{1/2}}{\partial q\ \partial q}-\frac{1}{n_{tot}}\frac{\partial^2 n_{tot}}{\partial q\ \partial q}\right)\cong D^{*1/2}\P_{(t)}+A_1\dot{q}\ , \qquad \text{(B.10)}$$

to

$$\ddot{q}_\sim=-\frac{A_1}{m}\dot{q}_\sim-\frac{1}{m}\frac{\partial\left(V_{(q)}+V_{qu}\right)}{\partial q_\sim}+\tilde{D}^{1/2}\P_{\sim(t)} \qquad \text{(B.11)}$$

where

$$\tilde{D}^{1/2}=\frac{D^{*1/2}}{m} \qquad \text{(B.12)}$$

The first order approximation (B.10) allows the Marcovian process to become self-consistent (independent by the dark matter evolution) reducing to and where $n_{(q,t)}=\int_{-\infty}^{+\infty}\mathcal{N}_{(q,p,t)}d^3p$ is defined by the Smolukowski equation (C.3) in appendix C, of the Marcovian process (2.3.7).

Moreover, for system with irrotational velocity field (that admits the action function $S$) such as $\frac{\partial S}{\partial q_i}=\dot{q}$, equation (2.3.7) can read

$$\dot{p}_i=-A_1\frac{\partial S}{\partial q_i}-\frac{\partial\left(V_{(q)}+V_{qu(n)}\right)}{\partial q_j}+m\tilde{D}^{1/2}\P_{(t)} \qquad \text{(B.13)}$$

that by posing

$$\frac{\partial\ \ S}{\partial q_\sim}=A_{1(q,t)}\frac{\partial S}{\partial q_\sim}, \qquad \text{(B.14)}$$

leads to

$$\ddot{q}_\sim=-\frac{1}{m}\frac{\partial\left(V_{(q)}+V_{qu}+\ \ S\right)}{\partial q_\sim}+\tilde{D}^{1/2}\P_{\sim(t)}, \qquad \text{(B.15)}$$

Besides, by comparing (B.15) with (2.3.4), it follows that

$$\overline{V}_{st} = \frac{\hbar^2}{2m}\left( \frac{1}{n^{1/2}} \frac{\partial^2 n^{1/2}}{\partial q\ \partial q} - < \frac{1}{n_{tot}{}^{1/2}} \frac{\partial^2 n_{tot}{}^{1/2}}{\partial q\ \partial q} > \right) \cong \quad S \qquad ( B.16)$$

and that

$$V_{st} = q^{1/2} q^{1/2} \overline{D}^{1/2} ¶_{\sim(t)} \qquad (B.17)$$

Finally it interesting to note that, for $\quad = m© \approx const$ the quantum hydrodynamic equation of motion leads to the quantum Brownian particle given by [44]

$$\ddot{q}_{\sim} = -©\dot{q}_{\sim} - \frac{1}{m}\frac{\partial\left(V_{(q)} + V_{qu}\right)}{\partial q_{\sim}} + ©D^{1/2}¶_{\sim(t)}. \qquad (B.18)$$

is recovered.

Unfortunately, the validity of (B.18) is not general since  is not constant.

This agree with the results given in ref. [45-46] that show that only in the case of linear harmonic oscillator, in contact with a classical heat bath, the friction  can be a constant. Besides, since in order to have the quantum decoupling with the environment (i.e., a classical super-system), the non-linear interaction is needed (see identity (3.5-6) of section 3), actually, the case $\quad =$ constant is never rigorously possible except for the deterministic limit of the canonical quantum mechanics with $\quad = 0$.

It can only approximaltely accepted for locally linear oscillators (non-linearly coupled to the environment) for which we can assume $\sim \approx 0$.

# Appendix C

## The environmental Marcovian noise in presence of the quatum potential

Since $n_{tot}$ is postulated by the approximation (2.8) the determination of  as well as of all the model is not complete.

Nevertheless, once infinitesimal dark matter fluctuations have broken the quantum coherence on the cosmological scale (i.e., $a_c \ll 10^{60} m$ that for barionic particles with mass $m \sim 10^{-(30 \div 27)} Kg$, it is enough $T \gg 10^{-120} \frac{4\hbar^2}{mk} {}^{\circ}K \sim 10^{-180 \circ} K$ ) and the resulting classical universe can be divided in subparts (the Newtonian limit of gravity is sufficiently weak force for satisfying condition (3.5)), we can define the super-system made up of the system and the environment.

At this stage, we can disergard the vacuum fluctuations associated to the dark matter (i.e., $¢n = 0$ ) and consider the Markovian process (2.3.7)

$$\dot{q}_i = \frac{p}{m}, \qquad (C.1)$$

$$\dot{p}_i = -\frac{\partial\left(V_{(q)} + < V_{qu(\tilde{n})} >\right)}{\partial q_j} + m©D^{1/2}¶_{(t)}, \qquad (C.2)$$

In presence of the quantum potential the evolution of the MDD $\tilde{n}_{(q,t)} = lim_{¢n\to 0}\, n_{tot}$ due to the stochastic motion equation (2.3.7)) depends by the exact sequence of the force inputs of the Marcovian noise.

On the other hand,, the probabilistic mass density (PMD) $\mathcal{N}_{(q,p,t)}$ of the Smoluchowski equation

$$P(q,p,q_0,p_0 \,|\, (t'+² - t_0), t_0) = \int_{-\infty}^{\infty} P(q,p,q',p' \,|²\, ,t')P(q',p',q_0,p_0 \,|\, t'-t_0,t_0) d^3q' \, d^3p'$$

(C.3)

for the Marcovian process (2.3.7) (where the PTF $P(x,z\,|²\,,t)$ represents the probability that an amount of the PMD) $\mathcal{N}_{(q,p,t)}$ at time $t$, in a temporal interval ², in a point $z = (q_0, p_0)$, transfers itself to the point $x = (q,p)$ [47]) is somehow indefinite since the quantum potential depends by the exact sequence of the inputs of the force noise.

Even if the connection between $\tilde{n}_{(q,t)}$ and $n(q,t)$ cannot be generally warranted, the approximation (B.10) that reads

$$-\frac{\hbar^2}{2m}\frac{\partial}{\partial q_{\sim}}\left( \frac{1}{n^{1/2}}\frac{\partial^2 n^{1/2}}{\partial q \ \partial q} - \frac{1}{\tilde{n}}\frac{\partial^2 \tilde{n}}{\partial q \ \partial q} \right) \cong m\bar{D}^{1/2}¶_{(t)} + A_1\dot{q} \quad , \qquad \text{(C.4)}$$

introduces the linkage between $\tilde{n}_{(q,t)}$ and

$$n_{(q,t)} = \int_{-\infty}^{+\infty} \mathcal{N}_{(q,p,t)} d^3p \qquad \text{(C.5)}$$

leading the motion equation

$$\dot{p}_i = -m©\dot{q}_{j(t)} - \frac{\partial\left(V_{(q)} + V_{qu_{(n)}}\right)}{\partial q_j} + m©D^{1/2}¶_{(t)} \qquad \text{(C.6)}$$

It is worth mentioning that the appplicability of (C.6) is not general but it is strongly subjected to the condition of being applied to small scale systems with $\mathcal{L} \ll {}^{a}{}_c$ that admit stationary states ($\bar{\dot{q}}_{\sim} = 0$) whose MDD is sufficently close to that of the deterministic eigenstates (i.e., small force noise amplitude) for which it is possible to assume that the collection of all MDD $\tilde{n}_{(q,p,t)}$ configurations will reproduce the PMD $n(q,t)$ such as

$$n_{(q)} = lim_{\Delta t\to\infty} \frac{1}{\Delta t} \int_{t-\frac{\Delta t}{2}}^{t\frac{\Delta t}{2}} \tilde{n} d² = \bar{\tilde{n}}_{(q)} \qquad \text{(C. 7)}$$

This assumption is at the basis of the relation (C,4) that expresses the connection betwen the PMD $n$ and the MDD $\tilde{n}$ (i.e., the information about $\tilde{n}$ can be obtained by knowing $n$ and $\dot{q}_{\sim}$ ).

Besides, if the system is sufficiently close to the deteministic limit of the quantum mechanics (for which $\mathcal{L} \ll {}^{a}{}_c$ (i.e., very small force noise amplitude) it is a sufficient condition) and it owns (irrotational [31]) stationary states (i.e., quasi-eigenstates) so that it is still quantum and. the action function $S$ (as integral of the momentum field) exists, we have that $\dot{q}_{\sim} = \frac{1}{m}\frac{\partial S}{\partial q_{\sim}}$ and $\tilde{n}$ can be defined by knowing $n$ and $S$ .

## C.1. The conservation equation of the Smolukowski equation in presence of the quantum potential

By using the method due to Pontryagin [47] the Smolukowski equation leads to the differential conservation equation for the PTF $P(q,z|^{2},t)$

$$\frac{\partial P_{(x,z|t,0)}}{\partial t} + \frac{\partial P_{(x,z|t,0)} \mathcal{V}_i}{\partial x_i} = 0 , \qquad \text{(C.1.1)}$$

where the current $J_i = P_{(x,z|t,0)} \mathcal{V}_i$ is given by the series of cumulants

$$P_{(x,z|t,0)} \mathcal{V}_i = P_{(x,z|t,0)} \bar{\dot{x}}_i - \frac{1}{2}\frac{\partial D_{im} P_{(x,z|t,0)}}{\partial x_m} + ... + \frac{1}{n!}\sum_{h=2}^{\infty} \frac{\partial^k C^{(k)}_{im........l} P_{(x,z|t,0)}}{\underbrace{\partial x_m ... \partial x_l}_{(2k-terms)}} \qquad \text{(C.1.2)}$$

where

$$C^{(k)}_{im........l} = lim_{2 \to 0} \frac{1}{2} \int_{-\infty}^{\infty} (y_i - x_i) \underbrace{(y_m - x_m)...(y_l - x_l)}_{(k-terms)} P_{(y,x|^{2},t)} d^{3h} y \qquad \text{(C.1.3)}$$

and where

$$\bar{\dot{x}}_{(q,p,t)} = \begin{pmatrix} \bar{\dot{q}} \\ \bar{\dot{p}} \end{pmatrix} = \begin{pmatrix} \dot{q} \\ -\frac{\partial \left( V_{(q)} + V_{qu} \right)}{\partial q_j} \end{pmatrix} , \qquad \text{(C.1.4)}$$

being

$$dx = d\begin{pmatrix} q \\ p \end{pmatrix} = \begin{pmatrix} \dot{q} \\ -\frac{\partial \left( V_{(q)} + V_{qu} \right)}{\partial q_j} \end{pmatrix} dt + \begin{pmatrix} 0 \\ m\left( \frac{< ´ (\dot{q}_{¥}), ´ (\dot{q}_{¡}) >_{(T)}}{a_c} \right)^{1/2} dW_{(t)} \end{pmatrix} . \qquad \text{(C.1.5)}$$

Moreover, for one particle problem or many decoupled particle system (e.g., linear oscillators)) it is possile the diagonal description

$$D_{im} = \begin{pmatrix} 0 & 0 \\ 0 & D_{p¥¡} \end{pmatrix} = \begin{pmatrix} 0 & 0 \\ 0 & m^2 \frac{< ´ (\dot{q}_{¥}), ´ (\dot{q}_{¡}) >_{(T)}}{a_c} ¢_{¥¡} \end{pmatrix}$$

$$= \begin{pmatrix} 0 & 0 \\ 0 & ¢_{¥¡}\, lim_{2 \to 0} \frac{1}{2} \int_{-\infty}^{\infty} \left( p_{¥} - \bar{p}_{¥} \right)\left( p_{¡} - \bar{p}_{¥} \right) P_{(p,\bar{p}|^{2},t)} d^3 p_{¡} \end{pmatrix} \qquad \text{(C.1.6)}$$

$$
C^{(k)}_{im....l} = \begin{bmatrix} 0 & 0 \\ 0 & C^{(k)}_{\underbrace{i\ ¥\,i\,...µ}_{k-indexes}} \end{bmatrix}_{2k\times 2k}
$$

$$
= \begin{pmatrix} 0 & 0 \\ 0 & lim_{²\to 0}\frac{1}{²}\int_{-\infty}^{\infty}\left(p_{¥}-\bar{p}_{¥}\right)\underbrace{\left(p_{i}-\bar{p}_{i}\right)...\left(p_{µ}-\bar{p}_{µ}\right)}_{(k-terms)}P_{(p,\bar{p}|²,t)}d^3p_{i}....d^3p_{µ} \end{pmatrix} \qquad \text{(C.1.7)}
$$

## C.2.1 The non-Gaussian PTF generated by the quantum potential

$$
\frac{\partial P_{(q,z|t,0)}}{\partial t}+\frac{\partial P_{(q,z|t,0)}\mathcal{V}_i}{\partial q_i}=0, \qquad (25) \qquad \text{(C.2.1.1)}
$$

where the current $J_i = P_{(x,z|t,0)}\mathcal{V}_i$ is given by the series of cumulants [47]

$$
P_{(q,z|t,0)}\mathcal{V}_i = P_{(q,z|t,0)} < \dot{q} >_i - \frac{1}{2}\frac{\partial D_{im}P_{(q,z|t,0)}}{\partial q_m}+...+\frac{1}{n!}\sum_{n=2}^{\infty}\frac{\partial^n C^{(n)}_{im_1........m_n}P_{(q,z|t,0)}}{\partial q_{m_1}....\partial q_{m_n}} \qquad \text{(C.2.1.2)}
$$

$$
\begin{aligned}
P_{(q,z|t,0)}\mathcal{V}_i = & -P_{(q,z|t,0)}\frac{1}{\partial q_i}\frac{\partial\left(V_{(q)}+V_{qu(P(q,q_0|t+²,t_0))}\right)}{\partial q_i} - \frac{1}{2}\frac{\partial D_{im}P_{(q,z|t,0)}}{\partial q_m} \\
& +..+\frac{1}{n!}\sum_{n=2}^{\infty}\frac{\partial^n C^{(n)}_{im_1........m_n}P_{(q,z|t,0)}}{\partial q_{m_1}....\partial q_{m_n}}
\end{aligned} \qquad \text{(C.2.1.3)}
$$

where

$$
C^{(n)}_{im_1........m_n} = lim_{²\to 0}\frac{1}{²}\int\left(y_i-q_i\right)\left(y_{m_1}-q_{m_1}\right)...\left(y_{m_n}-q_{m_n}\right)P_{(y,q|²,t)}d^{3h}y \qquad \text{(C.2.1.4)}
$$

that owns an infinite number of terms due to the presence of the quantum potential.

If on one hand, the continuity of the Hamiltonian potential warrants that velocities $\frac{\left(y_i-q_i\right)}{²}$ are finite and $C^{(n)}_{im_1........m_n}\to 0$ on very short time increment, on the other hand, since the quantum potential depends by the derivatives of $n_{(q,t)}$, it can lead to very high values of force also in the limit of very short time increment so that very far away points $\left(y_i-q_i\right)$ can contribute to the probability transition function $P(y,q\,|²,t)$ and the cumulants higher than two cannot be disregarded in (C.2.1.3).

Thence, being the cumulants higher than two non-vanishing, the PTF $P(q,z\,|²,t)$ is not Gaussian and. and equation (C.1.1) does not reduce to the FPE.

## C.3. The motion equation for the spatial densities

By integrating over the momenta, the conservation equation

$$\frac{\partial \mathcal{N}(q,p,t)}{\partial t}+\frac{\partial \mathcal{N}(q,p,t)\mathcal{V}_i}{\partial x_i}$$

$$=\frac{\partial \mathcal{N}_{(q,p,t)}}{\partial t}+\frac{\partial \mathcal{N}_{(q,p,t)}\bar{\dot{x}}_i}{\partial x_i}+\frac{\partial\left(\frac{1}{2}\frac{\partial C^{(1)}_{im}\mathcal{N}_{(q,p,t)}}{\partial x_m}+...+\frac{1}{n!}\sum_{n=2}^{\infty}\frac{\partial^n C^{(n)}_{im........l}\mathcal{N}_{(q,p,t)}}{\partial x_m....\partial x_l}\right)}{\partial x_i}$$

$$=\frac{\partial \mathcal{N}}{\partial t}-\frac{\partial \mathcal{N}\bar{\dot{q}}_{\sim}}{\partial q_{\sim}}-\frac{\partial \mathcal{N}\bar{\dot{p}}}{\partial p}+\frac{\partial\left(\frac{1}{2}\frac{\partial C^{(1)}_{im}\mathcal{N}_{(q,p,t)}}{\partial x_m}+...+\frac{1}{n!}\sum_{n=2}^{\infty}\frac{\partial^n C^{(n)}_{im........l}\mathcal{N}_{(q,p,t)}}{\underbrace{\partial x_m....\partial x_l}_{n-terms}}\right)}{\partial x_i}=0 \tag{C.3.1}$$

$$\mathcal{V}_i=\begin{pmatrix}\dot{\mathcal{Q}}\\ \dot{\mathcal{P}}\end{pmatrix}=\bar{\dot{x}}_i-\frac{1}{2P_{(x,z|t,0)}}\frac{\partial D_{im}P_{(x,z|t,0)}}{\partial x_m}+...+\frac{1}{n!\,P_{(x,z|t,0)}}\sum_{h=2}^{\infty}\frac{\partial^k C^{(k)}_{im.........l}P_{(x,z|t,0)}}{\underbrace{\partial x_m...\partial x_l}_{(2k-terms)}} \tag{C.3.2}$$

$$\bar{\dot{x}}_i=\begin{pmatrix}\bar{\dot{q}}\\ \bar{\dot{p}}=m\bar{\ddot{q}}_{\sim(t)}\end{pmatrix}=\begin{pmatrix}\dot{q}\\ -\frac{\partial\left(V_{(q)}+V_{qu}\right)}{\partial q_{\sim}}\end{pmatrix} \tag{C.3.3}$$

$$\dot{x}_i=\begin{pmatrix}\dot{q}\\ \dot{p}=m\ddot{q}_{\sim(t)}\end{pmatrix}=\begin{pmatrix}p\ \ /m\\ -\frac{\partial\left(V_{(q)}+V_{qu}\right)}{\partial q_{\sim}}+D_p^{\ 1/2}\P_{\sim(t)}\end{pmatrix} \tag{C.3.4}$$

where

$$n_{(q,t)}=\int \mathcal{N}(q,p,t)d^{3h}p\,, \tag{C.3.5}$$

$$V_{qu}=-\frac{\hbar^2}{2m}\frac{1}{n^{1/2}}\frac{\partial^2 n^{1/2}}{\partial q_{¥}\partial q_{¥}} \tag{C.3.6}$$

we obtain

$$\frac{\partial\int \mathcal{N}d^{3h}p}{\partial t}-\frac{\partial\int \mathcal{N}\dot{q}_{\sim}d^{3h}p}{\partial q_{\sim}}-<\dot{p}>_{(q,t)}\int_{-\infty}^{\infty}\frac{\partial \mathcal{N}}{\partial p}d^{3h}p$$

$$+\int\frac{\partial\left(\frac{1}{2}\frac{\partial C^{(1)}_{im}\mathcal{N}_{(q,p,t)}}{\partial x_m}+...+\frac{1}{n!}\sum_{n=2}^{\infty}\frac{\partial^n C^{(n)}_{im........l}\mathcal{N}_{(q,p,t)}}{\underbrace{\partial x_m....\partial x_l}_{n-terms}}\right)}{\partial x_i}d^{3h}p=0 \tag{C.3.7}$$

that with the condition $lim_{p\to\infty}\,\mathcal{N}_{(q,p,t)} = 0$ and by posing

$$< \dot{q}_{\sim} >= \frac{\int \mathcal{N}\dot{q}_{\sim}\, d^{3h}p}{\int \mathcal{N} d^{3h}p} \qquad \text{(C.3.8)}$$

leads to

$$\frac{\partial n}{\partial t} - \frac{\partial n < \dot{q}_{\sim} >}{\partial q_{\sim}} = -\int \frac{\partial\left(\frac{1}{2}\frac{\partial C^{(1)}_{im}\mathcal{N}_{(q,p,t)}}{\partial x_m} + \ldots + \frac{1}{n!}\sum_{n=2}^{\infty}\frac{\partial^n C^{(n)}_{im........l}\mathcal{N}_{(q,p,t)}}{\underbrace{\partial x_m \ldots\ldots \partial x_l}_{n-terms}}\right)}{\partial x_i} d^{3h}p \qquad \text{(C.3.9)}$$

where $C^{(1)}_{im} = p\begin{pmatrix} 0 & 0 \\ 0 & \tilde{D}¢_{\sim} \end{pmatrix}$ and where

$$-\int \frac{\partial\left(\frac{1}{2}\frac{\partial C^{(1)}_{im}\mathcal{N}_{(q,p,t)}}{\partial x_m} + \ldots + \frac{1}{n!}\sum_{n=2}^{\infty}\frac{\partial^n C^{(n)}_{im........l}\mathcal{N}_{(q,p,t)}}{\underbrace{\partial x_m \ldots\ldots \partial x_l}_{n-terms}}\right)}{\partial x_i} d^{3h}p$$

$$= \int \begin{pmatrix} 0 \\ \frac{\partial\left(\frac{1}{2}\frac{\partial \tilde{D}¢_{\sim ¡}\mathcal{N}_{(q,p,t)}}{\partial p_{¡}} + \ldots + \frac{1}{n!}\sum_{h=2}^{\infty}\frac{\partial^k C^{(k)}_{\sim ¡ ........£}\mathcal{N}_{(q,p,t)}}{\underbrace{\partial p_{¡} \ldots \partial p_{£}}_{(k-terms)}}\right)}{\partial p_{\sim}} \end{pmatrix} d^{3h}p \qquad \text{(C.3.10)}$$

$$Q_{diss_{(q,t)}} = \int \begin{pmatrix} 0 \\ \frac{1}{2}\frac{\partial \tilde{D}\mathcal{N}_{(q,p,t)}}{\partial p_{\sim}} + \ldots + \frac{1}{n!}\sum_{h=2}^{\infty}\frac{\partial^k C^{(k)}_{\sim ¡ ........£}\mathcal{N}_{(q,p,t)}}{\underbrace{\partial p_{¡} \ldots \partial p_{£}}_{(k-terms)}} \end{pmatrix} d^{3h}p \qquad \text{(C.3.11)}$$

gives the compressibility of the mass density distribution that is linked to the generation of entropy and quantum dissipation. Thence, equation (C.3.9) can read

$$\frac{\partial n}{\partial t} = \frac{\partial n}{\partial q_{\sim}} < \dot{q}_{\sim} > + n\frac{\partial}{\partial q_{\sim}} < \dot{q}_{\sim} > + Q_{diss_{(q,t)}} \qquad \text{(C.3.12)}$$